\documentclass[12pt]{JHEP3}
\usepackage{amsfonts}
\usepackage{amssymb}
\usepackage{lscape}
\usepackage{cite}
\usepackage{epsfig}
\usepackage{multirow}
\usepackage{longtable}
\usepackage[all]{xy}
\usepackage{amsmath}
\author{}

\newcommand{\be}{\begin{equation}}
\newcommand{\ee}{\end{equation}}
\newcommand{\ba}{\begin{array}}
\newcommand{\ea}{\end{array}}
\newcommand{\bea}{\begin{eqnarray}}
\newcommand{\eea}{\end{eqnarray}}

\newcommand{\ov}{\overline}
\def\IR{\relax{\rm I\kern-.18em R}}

\def\IP{\relax{\rm I\kern-.18em P}}
\def\inbar{\vrule height1.5ex width.4pt depth0pt}
\def\IC{\relax\,\hbox{$\inbar\kern-.3em{\rm C}$}}

\def\K3{{\bf K3}}

\def\ov{\overline}

\def\n2d{\cN_{V^*}^{\otimes 2}}

\def\IC{\mathbb{C}}

\def\IR{\mathbb{R}}

\def\IP{\mathbb{P}}

\def\cN{{\mathcal N}}

\def\nn{\nonumber}

\def\Tr{{\mathop {\rm Tr}}}
\title{Light stringy states}
\author{
Pascal Anastasopoulos$^{1}$\footnote{pascal@hep.itp.tuwien.ac.at},~
Massimo Bianchi$^{2}$\footnote{Massimo.Bianchi@roma2.infn.it},~
Robert Richter$^{2,3}$\footnote{rrichter@roma2.infn.it}~\\
$^1$ Technische Univ. Wien Inst. f\"ur Theoretische Physik, A-1040 Vienna, Austria\\
$^2$ Dipartimento di Fisica \& Sezione I.N.F.N., Universit\`a di Roma ``Tor Vergata'', \\
Via della Ricerca Scientica, 00133 Roma, Italy \\
$^3$ II. Institut f\"ur Theoretische Physik, Hamburg University, Germany\\
}
\date{}

\maketitle
\abstract{
We carefully study the spectrum of open strings localized at the intersections of D6-branes
and identify the lowest massive `twisted' states and their vertex operators, paying particular attention to the signs of the intersection angles. We argue that the masses of the lightest states scale as $M^2_{\theta} \approx \theta M_s^2$ and can thus be parametrically smaller than the string scale. Relying on previous analyses, we compute scattering amplitudes of massless `twisted' open strings and study their factorization, confirming the presence of the light massive states as sub-dominant poles in one of the channels.
}
\preprint{
TUW-11-23\\
ROM2F/2011/14\\
ZMP-HH/11-16}
 \clearpage
\begin{document}

\section{Introduction}

Vacuum configurations with open unoriented strings have attracted a lot of attention in the past few years for their remarkable phenomenological properties\cite{Blumenhagen:2005mu,Blumenhagen:2006ci,Marchesano:2007de,Bianchi:2009va}\footnote{For reviews on phenomenological implications of D-instantons in this context, see \cite{Blumenhagen:2009qh,Bianchi:2009ij,Cvetic:2011vz} }. One of the peculiar features is the possibility of accommodating large extra dimensions giving rise to a significantly lower string scale, even of a few $TeV$ \cite{ArkaniHamed:1998rs,Antoniadis:1997zg,Antoniadis:1998ig}. Scenarios of these kinds may circumvent the hierarchy problem, but also allow for stringy signatures that can be observed at LHC \cite{Dudas:1999gz,Accomando:1999sj,Cullen:2000ef,Kiritsis:2002aj,
Burikham:2004su,Chialva:2005gt,Bianchi:2006nf,Anchordoqui:2007da,Anastasopoulos:2008jt,Anchordoqui:2008ac,Dong:2010jt,Feng:2010yx,Anchordoqui:2010zs,Feng:2011qc}.

Recently, in a series of papers \cite{Lust:2008qc,Anchordoqui:2009mm,Lust:2009pz,Anchordoqui:2009ja} the authors study tree-level string scattering amplitudes containing at most two chiral fermions. They show that these amplitudes exhibit a universal behaviour independently of the specifics of the compactification, which gives their results a predictive power. The observed poles correspond to the exchanges of Regge excitations of the standard model gauge bosons, whose masses scale with the string mass $M_s$. On the other hand there exists a tower of stringy excitations of the chiral fermions and their superpartners localized at the intersections of two stacks of D-branes. Their masses depend on the string mass $M_s$ and the intersection angle $\theta$ and thus can be significantly lighter than the Regge excitations of the gauge bosons.

A large subclass of semi-realistic global D-brane constructions exhibit small intersection angles between two stacks of D-branes and thus allow for light stringy states. {\it A priori} the widths of the angles depend on the wrapping numbers of the intersecting branes and on the moduli of the compactification, associated to closed-string excitations. Playing with both discrete and continuous degrees of freedom it is possible to lower the threshold for the production of these states well below the string scale $M_s \approx \sqrt{T_s}$. %Similar considerations apply to (generalized) Kaluza-Klein excitations, that we will not delve upon very much here.
Aim of the present work is the investigation of massive, but potentially very light, open string states. We analyze in detail a configuration of intersecting D-branes, discuss the states arising at such an intersection beyond the massless level. Moreover, we give a detailed description for the construction of their vertex operators, which crucially depends on the signs of the intersection angles.

Equipped with the vertex operators for arbitrary intersection angles we compute the four point amplitude containing four fermions. We investigate various limits of this amplitude and show that the most dominant poles correspond to the exchanges of the light stringy states. While the signals of such light stringy states at colliders could be not so easy to recognize and discriminate from other kinds of Physics Beyond the Standard Model the amplitude also exhibits signatures of higher spin exchanges, whose origin is purely stringy and whose masses do not vanish for small angles. Thus signatures of light stringy states may provide a first step towards evidence for string theory.

The presentation will be organized as follows. In section \ref{sec vertex operators}, we present a dictionary between massless or massive states localized at two intersecting D-brane stacks and their corresponding vertex operator.
%discuss a local configuration of two intersecting D-brane stacks, analyze the states localized at such intersection and eventually display their corresponding masses and vertex operators. 
In section \ref{sec amplitude} we
 will compute some relevant scattering amplitudes at tree-level (disk) and expose the massive poles associated to massive, but light open strings. In section \ref{sec concl} we will conclude. The appendix \ref{app bosonic twist fileds} provides the definitions of the bosonic twist fields appearing in the vertex operators, while in appendix \ref{app massive states} we apply the state - vertex operator dictionary laid out in section \ref{sec vertex operators} to some  particular massive states localized at the intersection of two D-branes. The appendices \ref{app correlators} and \ref{app hypergeometric} provide some technical details necessary for the computation and analysis of the considered amplitude.

\section{Quantization of strings localized at D-brane intersections
\label{sec vertex operators}}
In this section we will analyze the states localized at the intersection of two stacks of D6-branes. We will derive a dictionary between states localized at such an intersection and their corresponding vertex operators%We will investigate under which conditions the states are massless and identify their corresponding vertex operators
\footnote{For a discussion of vertex operators for massless states at arbitrary intersection angles, see \cite{Bertolini:2005qh, Cvetic:2006iz}. For an analysis of instantonic modes at the intersection of  D-instanton and D-brane at arbitrary angles, see \cite{Cvetic:2009mt}. Vertex operators of massive states in heterotic compactifications are discussed in \cite{Bianchi:2010es}.}.
Let us start by solving the equations of motion for an open string stretched between two D-brane stacks intersecting at an angle $\pi\theta$ in the $({{X}}, {{Y}})$ plane. The bosonic coordinates have to fulfil the boundary conditions
\cite{Berkooz:1996km,Arfaei:1996rg,Abel:2003vv}
\begin{align}
\begin{gathered}
\partial_{\sigma} {{X}}(\tau,0) = 0 = {{Y}} (\tau,0) \\
\partial_{\sigma} {{X}} (\tau,\pi) + \tan{(\pi{{\theta}})} \,\,\partial_{\sigma} {{Y}}(\tau,\pi)=0
\\
 {{Y}} (\tau,\pi)- \tan{(\pi{{\theta}})} \,\, {{X}} (\tau,\pi)=0\,\,.
\end{gathered}
\label{eq boundary conditions}
\end{align}
It proves convenient to introduce complex coordinates $Z^{I}=X^{I} + \mathrm{i}
Y^{I}$ with $I=1,2,3$ for the internal (compactified) directions.
Given these boundary conditions for each $X$ and $Y$, one can deduce the mode expansions
for each $\partial Z$ and $\partial \ov Z$ that read (after applying the doubling trick) 
\begin{align}
\partial Z (z) = \sum_{n}
\alpha_{n-\theta} \,\,z^{-n+\theta-1} \qquad \qquad  \partial \ov Z (z) = \sum_{n}
\alpha_{n+\theta} \,\,z^{-n-\theta-1}\,\,.
\end{align}
%\textbf{\emph{Here seems to be a minus sign problem, however I think one can perform a redefinition of $\alpha^I_{n+\theta_I}$ that will take care of that. I will try to verify that later.}}
Upon quantization the only non-vanishing commutators are
\begin{align*}
[\alpha_{n \pm \theta}, \alpha_{m \mp \theta} ] &=  (m \pm \theta)
\,\delta_{n+m}\,\,.
\end{align*}
World-sheet supersymmetry
$\delta X= \ov \epsilon \psi$
leads to the same modding for the complexified world-sheet fermions. One obtains (again after using the doubling trick)
\begin{align}
\Psi(z)= \sum_{r\in\mathbb{Z}+\nu} \,\, \psi_{r-\theta}\,
z^{-r-\frac{1}{2}+\theta} \qquad \ov{\Psi}(z)=
\sum_{r\in\mathbb{Z}+\nu} \,\, \psi_{r+\theta}\,
\bar{z}^{-r-\frac{1}{2}-\theta}\,\,,
\label{modeexpansionintersecting fermionic}
\end{align}
where $\nu$ is $\frac{1}{2}$ and $0$ for the NS-sector and R-sector, respectively. 
Upon quantization the only non-vanishing anti-commutator are
\begin{align}
\{\psi_{m-\theta}, \psi_{n+\theta} \}=\delta_{m,n}\,\,.
\label{eq anti-comm}
\end{align}

In the following we present a prescription that gives the vertex operator corresponding  to any state localized at an intersection of two D-branes.  
To this end we need to properly define the ground-state and identify the annihilation and creation operators. Equipped with the proper ground state definition we derive the OPE's of the conformal fields $\partial Z$, $\partial \ov Z$, $\Psi$ and $\ov \Psi$ with the vacua and excitations thereof. With their knowledge one is able to write down the vertex operator corresponding to any state, be it massless or massive.

\subsection{NS-sector \label{sec NS sector}}
Let us start with the NS sector that describes space-time bosons restricting for the moment our attention onto just one complex dimension. The definition of the ground-state crucially depends on whether the intersection angles are positive or negative.  For a positive intersection angle the ground-state $|\,\theta_I\,\rangle_{NS}$ is given by
\begin{equation}
\begin{aligned}
\alpha_{m-\theta} |\,\theta\,\rangle_{NS} &=0   \qquad m \geq 1 \qquad
\qquad
\psi_{r-\theta} |\,\theta\,\rangle_{NS} =0 \qquad r \geq \frac{1}{2}\\
 \alpha_{m+\theta} |\,\theta\,\rangle_{NS} &=0 \qquad m \geq 0\qquad
\qquad \psi_{r+\theta} |\,\theta\,\rangle_{NS} =0 \qquad r \geq
\frac{1}{2}\,\,.
\end{aligned}
\end{equation}
whereas for a negative intersection angle it is defined as

\begin{equation}
\begin{aligned}
\alpha_{m-\theta} |\,\theta\,\rangle_{NS} &=0   \qquad m \geq 0 \qquad
\qquad
\psi_{r-\theta} |\,\theta\,\rangle_{NS} =0 \qquad r \geq \frac{1}{2} \\
\alpha_{m+\theta} |\,\theta\,\rangle_{NS} &=0 \qquad m \geq 1 \qquad \qquad
 \psi_{r+\theta} |\,\theta\,\rangle_{NS} =0 \qquad r \geq \frac{1}{2}\,\,.
 \label{eq def. of vacuum}
\end{aligned}
\end{equation}

Due to the non-trivial intersection angles the vertex operators describing the states under consideration involve bosonic and fermionic twist fields accounting for the boundary conditions \eqref{eq boundary conditions}. In order to properly identify these twist fields we determine the action of the conformal fields $\Psi$, $\ov \Psi$, $\partial Z$ and $\partial \ov Z$ on the ground-state $|\,\theta\,\rangle_{NS}$ and excitations (fermionic and bosonic ones) thereof.

We start by investigating the ground-state $|\,\theta\,\rangle_{NS}$, with positive intersection angle $\theta$, that can be identified with  $s^+_{\theta}(0) \,\sigma^+_{\theta}(0)|\,0\,\rangle^u_{NS}$, where $s^+_{\theta}$, $\sigma^+_{\theta}$ denote the fermionic and bosonic twist fields, respectively, and $|\,0\,\rangle^u_{NS}$ is the untwisted vacuum. 
Acting with the bosonic conformal fields $\partial Z$ and $ \partial \ov Z$ on the  the ground-state $| \, \theta \,\rangle_{NS}$ we obtain
\begin{align*} 
& \partial Z(z) | \, \theta \,\rangle_{NS}=\sum_{n\in \mathbb{Z}}
\alpha_{n-\theta} z^{-n+\theta-1}  | \, \theta \,\rangle_{NS} 
\rightarrow  z^{\theta-1} \,  \alpha_{-\theta}  | \, \theta \,\rangle_{NS} = z^{\theta-1}  s^+_{\theta}(0)  \tau^+_{\theta}(0)| \, 0 \,\rangle^u_{NS} \\ 
& \partial \ov Z(z) | \, \theta \,\rangle_{NS} = \sum_{n\in \mathbb{Z}}
\alpha_{n+\theta} z^{-n-\theta-1}  | \, \theta \,\rangle_{NS} 
\rightarrow  z^{-\theta} \,  \alpha_{-1+\theta}  | \, \theta \,\rangle_{NS} = z^{-\theta}  s^+_{\theta}(0)  \widetilde{\tau}^+_{\theta}(0)| \, 0 \,\rangle^u_{NS} \,\,.
\end{align*}
Here $\tau^+_{\theta}$ and $\widetilde{\tau}^+_{\theta}$ denote excited twist fields with conformal dimensions $h_{\tau^+_{\theta}} = \frac{1}{2}\theta (3-\theta) $ and $h_{\widetilde{\tau}^+_{\theta}} = \frac{1}{2}(1-\theta)(2+\theta)$, respectively. Analogously acting with $\Psi$ and $\ov \Psi $ on the twisted vacuum $|\,\theta \,\rangle_{NS}$ gives
\begin{align*}
\Psi (z) \, |\,\theta\,\rangle_{NS} &= \sum_{r\in \mathbb{Z}+\frac{1}{2}}
z^{-r-\frac{1}{2}+\theta} \psi_{r-\theta} |\,\theta \rangle_{NS}
\, \rightarrow \,\,z^{\theta } \psi_{-\frac{1}{2}-\theta } |\,\theta \,\rangle_{NS} =
 z^{\theta } \,
\tilde{t}^+_{\theta }(0) \,\sigma^+_{\theta }(0) |\,0\,\rangle^u_{NS}\,\,\\ 
\ov{\Psi} (z)\, |\,\theta\,\rangle_{NS}&= \sum_{r\in \mathbb{Z}+\frac{1}{2}}
z^{-r-\frac{1}{2}+\theta } \psi_{r+\theta} |\,\theta \rangle_{NS} \, \rightarrow \,\,z^{-\theta} \, \psi_{-\frac{1}{2}+\theta}  |\,\theta\,\rangle_{NS} =z^{-\theta} \,
t_{\theta}^+(0) \, \sigma^+_{\theta}(0) |\,0\,\rangle^u_{NS} \,\,,
\end{align*}
where $\tilde{t}^+_{\theta}$ and $t_{\theta}^+$ denote excited fermionic twist fields with conformal dimension $h_{\tilde{t}^+_{\theta}}=\frac{1}{2} \left( 1+\theta  \right)^2$ and $h_{t^+_{\theta}}=\frac{1}{2} \left( 1-\theta \right)^2$, respectively. The fermionic conformal fields allow for a bosonization  which then take the form
\begin{align}
& \hspace{25mm} \Psi(z) = e^{iH(z)} \qquad \qquad \qquad  \ov \Psi(z) = e^{-iH(z)} \\
&s^+_{\theta} (z) = e^{i\theta H (z)} \qquad \qquad  t^+_{\theta} (z) = e^{i( \theta+1) H (z)} \qquad \qquad \tilde{t}^+_{\theta} (z) = e^{i (\theta-1) H (z)} \nn\,\,.
\end{align}
Analogously one can apply the same procedure to fermionic and bosonic excitations of the vacuum as well as the ground state for negative intersection angle, defined in \eqref{eq def. of vacuum} and excitations thereof. We display our findings in the table \ref{table Excitations NS sector}.
\begin{table}[h] \centering
\begin{tabular}{| l | l || l | l |}
\hline
\multicolumn{2}{|c||}{Positive angles}&
\multicolumn{2}{|c|}{Negative angles} 
\\
\hline
\hline
state & vertex operator &state & vertex operator\\ \hline \hline
$| \, \theta\, \rangle_{NS} $ & 
$  e^{i \theta H(z)} \sigma^+_{\theta}(z) $ & $| \, \theta \, \rangle_{NS}  $& 
$  e^{i \theta H(z)} \sigma^-_{\theta}(z) $\\ \hline
$\alpha_{-\theta} | \, \theta\, \rangle_{NS}$ & 
$  e^{i \theta H_(z) }\tau^+_{\theta}(z)$ & 
$\alpha_{\theta} | \, \theta\, \rangle$ & 
$  e^{i \theta H(z) }\widetilde\tau^-_{\theta}(z) $\\ \hline
$\left(\alpha_{-\theta} \right)^2| \, \theta\, \rangle_{NS}$ & 
$  e^{i \theta H(z)} \omega^+_{\theta}(z) $&
$ \left(\alpha_{\theta} \right)^2| \, \theta\, \rangle_{NS}$ & 
$  e^{i \theta H(z)} \widetilde\omega^-_{\theta}(z)$ \\ \hline
$ \psi_{-\frac{1}{2} + \theta}| \, \theta \, \rangle_{NS} $ & 
$ e^{i \left(\theta-1\right)H(z)} \sigma^+_{\theta}(z)$ & 
$\psi_{-\frac{1}{2} - \theta}| \, \theta \, \rangle_{NS} $ & 
$ e^{i \left(\theta+1\right)H (z)} \sigma^-_{\theta}(z) $ \\ \hline
$ \alpha_{-\theta}\, \psi_{-\frac{1}{2} + \theta} | \, \theta\, \rangle_{NS}$ & 
$  e^{i \left(\theta -1\right)H(z) }\tau^+_{\theta}(z) $ &
$ \alpha_{\theta}\, \psi_{-\frac{1}{2} - \theta} | \, \theta\, \rangle_{NS}$ & 
$  e^{i \left(\theta +1\right)H(z) }\widetilde\tau^-_{\theta}(z) $ \\ \hline
$ \left(\alpha_{-\theta} \right)^2 \psi_{-\frac{1}{2} + \theta} | \, \theta\, \rangle_{NS}$ & 
$  e^{i\left(\theta-1\right) H(z)} \omega^+_{\theta}(z) $& 
$ \left(\alpha_{\theta} \right)^2 \psi_{-\frac{1}{2} - \theta} | \, \theta\, \rangle_{NS}$ & 
$  e^{i\left(\theta+1\right) H(z)} \widetilde\omega^-_{\theta}(z)  $\\ \hline
$ \alpha_{-1+\theta} | \, \theta\, \rangle_{NS}$ & 
$  e^{i \theta H(z) }\widetilde{\tau}^+_{\theta}(z) $ & 
$\alpha_{-1-\theta} | \, \theta\, \rangle_{NS}$ & 
$  e^{i \theta H(z) }{\tau}^-_{\theta}(z) $ \\ \hline
$ \alpha_{-1+\theta} \psi_{-\frac{1}{2} + \theta} | \, \theta\, \rangle_{NS}$ &
$  e^{i \left(\theta -1\right)H(z) } \widetilde{\tau}^+_{\theta}(z) $ & 
$\alpha_{-1-\theta} \psi_{-\frac{1}{2} -\theta} | \, \theta\, \rangle_{NS}$ & 
$  e^{i \left(\theta +1\right)H(z) } {\tau}^-_{\theta}(z) $\\ \hline
\end{tabular}\nn
\caption{\small {Excitations and their corresponding vertex operator part for the NS-sector.}} % \vspace{3mm}
\label{table Excitations NS sector}
\end{table}

Here we give the bosonized form of the fermionic twist operators. In appendix \ref{app bosonic twist fileds} we properly define the bosonic twist fields by displaying their OPE's with $\partial Z$ and $\partial \ov Z$.     

\subsection{R-sector}

Let us turn to the R-sector, whose bosonic part is exactly the same as for the NS-sector. Thus it is sufficient to study the fermionic part. 
The mode expansion of $\Psi$ and $\ov \Psi$ are similar to the expansions in the  NS sector however the sum is over integers and not half-integers (see eq. \eqref{modeexpansionintersecting fermionic}).
Again the definition of the ground state crucially depends on whether the intersection angle is positive or negative. For positive intersection angle one has
\begin{equation}
\begin{aligned}
\alpha_{m-\theta} |\,\theta\,\rangle_R &=0   \qquad m \geq 1 \qquad
\qquad
\psi_{r-\theta} |\,\theta\,\rangle_R =0 \qquad r \geq 1\\
 \alpha_{m+\theta} |\,\theta\,\rangle_R &=0 \qquad m \geq 0\qquad
\qquad \psi_{r+\theta} |\,\theta\,\rangle_R =0 \qquad r \geq
0\,\,.
\end{aligned}
\end{equation}
whereas for a negative intersection angle one defines
\begin{equation}
\begin{aligned}
\alpha_{m-\theta} |\,\theta\,\rangle_R &=0   \qquad m \geq 0 \qquad
\qquad
\psi_{r-\theta} |\,\theta\,\rangle_R =0 \qquad r \geq 0\\
\alpha_{m+\theta} |\,\theta\,\rangle_R &=0 \qquad m \geq 1 \qquad \qquad
 \psi_{r+\theta} |\,\theta\,\rangle_R =0 \qquad r \geq 1\,\,.
 \label{eq def. of vacuum R}
\end{aligned}
\end{equation}
As one can easily see the bosonic part of the R-sector behaves similar as in the NS-sector. On the other hand due to the fact that the mode expansion of $\Psi$ and $\ov \Psi$ in the R-sector is over integers rather than half-integers the fermionic twist operators will take a different form from the ones in the NS-sector. Applying the same procedure as in the NS sector to obtain the necessary OPE's we get  the vacuum  $|\, \theta\,\rangle_R$. In case of positive intersection angle, it can be identified with $S^+_{\theta}(0) \sigma_{\theta}(0) | \, 0 \,\rangle^u_{R}$ {\it viz.}
\begin{align*}
\Psi (z)\, |\,\theta\,\rangle_R  =\sum_{n \in \mathbb{Z}} \,\, \psi_{n-\theta}\,
z^{-n-\frac{1}{2}+\theta}   |\, \theta\,\rangle_R ~~~  
 &\longrightarrow %~~~  z^{-\frac{1}{2}+\theta_I}  \,  \psi_{-\theta_I}^I |\,\theta_{I}\,\rangle = 
 z^{-\frac{1}{2}+\theta} T^+_{\theta}(0)\,\sigma_{\theta}(0) |\,0\,\rangle^u_R = |\, \theta\,\rangle_R\\
\ov \Psi(z) \, |\, \theta\,\rangle_R  =\sum_n \,\, \psi_{n+\theta}\,
z^{-n-\frac{1}{2}-\theta}   |\, \theta\,\rangle_R ~~~   
&\longrightarrow %~~~  z^{\frac{1}{2}-\theta_I}  \,  \psi_{-1+\theta_I}^I |\, \theta_I\,\rangle = 
z^{\frac{1}{2}-\theta} \widetilde{T}^+_{\theta}(0)\, \sigma_{\theta}(0) |\, 0\,\rangle^u_R = |\, \theta\,\rangle_R\,\,.
\end{align*}
Here $T^+_{\theta}$ and $ \widetilde{T}^+_{\theta}$ denote excited twist fields that can be bosonized
\begin{align}
S^+_{\theta}(z) =e^{i\left( \theta -\frac{1}{2}\right) H(z)}  \qquad T^+_{\theta}(z)= e^{i\left( {\theta}+\frac{1}{2}\right) H(z)} \qquad \widetilde{T}^+_{{\theta}} (z)=e^{i\left( {\theta} -\frac{3}{2}\right) H(z)} 
\end{align}

Analogously we can derive the vertex operator corresponding to any excitation. The definitions of the bosonic twist operators, namely their OPE's with the conformal fields $\partial Z$ and $\partial \ov Z$ are given in the appendix \ref{app bosonic twist fileds}. We summarize our findings in the table below, where the fermionic twists are given in the bosonized form as in the NS-sector.
\begin{table}[h] \centering
\begin{tabular}{| l | l || l | l |}
\hline
\multicolumn{2}{|c||}{Positive angles}&
\multicolumn{2}{|c|}{Negative angles} 
\\
\hline
\hline
state & vertex operator &state & vertex operator\\ \hline \hline
$ | \, \theta \, \rangle_R $ & 
$  e^{i \left(\theta-\frac{1}{2}\right) H(z)} \sigma^+_{\theta}(z) $ & 
$ | \, \theta \, \rangle_R $ & 
$  e^{i \left(\frac{1}{2}-\theta\right) H(z) } \sigma^-_{\theta}(z) $\\ \hline
$ \alpha_{-\theta} | \, \theta\, \rangle_R $ & 
$  e^{i \left(\theta-\frac{1}{2}\right) H(z) }\tau^+_{\theta}(z) $ & 
$ \alpha_{\theta} | \, \theta\, \rangle_R $ & 
$  e^{i \left(\frac{1}{2}-\theta\right) H(z) }\widetilde\tau^-_{\theta}(z) $ \\ \hline
%\left(\alpha_{-\theta_I} \right)^2| \, \theta_I\, \rangle &   e^{i \theta_I H_I(z)} \omega^+_{\theta_I}(z) &\left(\alpha_{\theta_I} \right)^2| \, \theta_I\, \rangle &   e^{i \theta_I H_I(z)} \widetilde\omega^-_{\theta_I}(z) \\ \hline
$ \psi_{- \theta}| \, \theta \, \rangle_R  $ &  
$  e^{i \left(\theta+\frac{1}{2}\right)H(z)} \sigma^+_{\theta}(z) $ & 
$ \psi_{ \theta}| \, \theta \, \rangle_R $ & 
$  e^{i \left(\theta -\frac{1}{2}\right)H(z)} \sigma^-_{\theta}(z) $ \\ \hline
$ \alpha_{-\theta}\, \psi_{-\theta} | \, \theta\, \rangle_R $ &  
$  e^{i \left(\theta +\frac{1}{2}\right) H(z) }\tau^+_{\theta}(z) $ &
$ \alpha_{\theta}\, \psi_{\theta} | \, \theta\, \rangle_R $ & 
$ e^{i \left(\theta -\frac{1}{2}\right)H(z) }\widetilde\tau^-_{\theta}(z) $ \\ \hline
%\left(\alpha_{-\theta_I} \right)^2 \psi_{-\frac{1}{2} + \theta} | \, \theta_I\, \rangle &   e^{i\left(\theta_I-1\right) H_I(z)} \omega^+_{\theta_I}(z)&
%\left(\alpha_{\theta_I} \right)^2 \psi_{-\frac{1}{2} - \theta} | \, \theta_I\, \rangle &   e^{i\left(\theta_I+1\right) H_I(z)} \widetilde\omega^-_{-\theta_I}(z)  \\ \hline
$ \psi_{-1+ \theta} | \, \theta\, \rangle_R $ & 
$ e^{i \left(\theta-\frac{3}{2}\right) H(z) }\sigma^+_{\theta}(z) $ & 
$ \psi_{-1- \theta}  | \, \theta\, \rangle_R $ & 
$ e^{i\left(\theta+\frac{3}{2} \right)H(z) } \sigma^-_{\theta}(z) $ \\ \hline
$ \alpha_{-\theta}\,  \psi_{-1 + \theta} | \, \theta\, \rangle_R $ ~~& 
$e^{i \left(\theta-\frac{3}{2}\right) H(z) }\tau^+_{\theta}(z) $ ~~&
$ \alpha_{\theta}\, \psi_{-1 -\theta} | \, \theta\, \rangle_R $ ~~&
$ e^{i (\left(\theta +\frac{3}{2}\right) H(z) } \widetilde{\tau}^-_{\theta}(z) $ ~~\\ \hline
\end{tabular}
\caption{\small {Excitations and their corresponding vertex operator part for the R-sector.}} % \vspace{3mm}
\label{table Excitations R sector}
\end{table}

% \vspace{-3mm}

\subsection{States and vertex operators 
\label{sec states NS}}
In the previous subsection we derived the necessary building blocks of the vertex operators. Here we will display  the vertex operators corresponding to specific states. Before turning to concrete examples we give the mass formula, which can be easily derived from the Virasoro operator  \cite{Aldazabal:2000dg} 
\begin{align}
M^2&=\left( \sum^2_{\mu=1} \left\{\sum_{n \epsilon \emph{Z} }:
\alpha^{\mu}_{-n} \, \alpha^{\mu}_{n}:+ \sum_{n \epsilon \emph{Z}}
n\,:\psi^{\mu}_{-n}\, \psi^{\mu}_{n}:\right\} \right. \\ &~~~ \left.+ \sum^3_{I=1}
\left\{\sum_{m\epsilon
 \emph{Z}}:\alpha^I_{-m+\theta_I}\alpha^I_{m-\theta_I}: +
\sum_{m\epsilon \emph{Z}+\nu}
(m-\theta_I):\psi^I_{-m+\theta_I}\psi^I_{m-\theta_I}:\right\}
+\epsilon^I_0\right) M^2_s \,\,. \nn
\end{align}
Here $\nu$ is $\frac{1}{2}$ and $0$ for the NS- and R-sector, respectively and the index $I$ denotes the internal dimension. The zero point energy $\epsilon^I_0$ of the $I$-th dimension can be computed by $\zeta$-function regularization to
$\epsilon^{I}_0= -\frac{1}{8}+\frac{1}{2}\, \theta_I$ ( $\epsilon^{I}_0= -\frac{1}{8}-\frac{1}{2}\, \theta_I$ ) for positive (negative) intersection angle for the the NS-sector and $\epsilon^{I}_0=0$ for the R-sector.

For supersymmetric intersections,  we are mosty  interested in, the three intersection angles have to satisfy the following condition
\begin{align}
\theta_1 + \theta_2 + \theta_3 = 0 \qquad  \text{mod} \,\,\,\, 2
\end{align}
which leaves the following two independent options
\begin{itemize}
 \item[$\bullet$]  $\theta_1, \theta_2, \theta_3 \geq 0 $ with $\sum_I \theta_I =2$
 \item[$\bullet$]  $\theta_1, \theta_2 \geq 0 $ and $ \theta_3 \leq  0 $ with $\sum_I \theta_I =0$\,\,.
\end{itemize}
%where the last two options are just the mirror of the first two and are therefore already covered by them. 
Below we will discuss these two setups in detail, we present the massless states in the NS- and R-sector, display their corresponding vertex operator and then turn to genuinely massive string states discuss their masses as well as their vertex operators. For a more complete list of massive states localized at the intersection of two D-branes we refer to the appendix \ref{app massive states}.

Finally, not all possible excitations correspond to physical states.  The GSO projection, ensuring modular invariance of the parent closed-string partition function, requires that a physical state in the NS-sector contains an odd number of fermionic excitations.

\subsubsection*{Only positive angles}

Let us start with the setup in which all intersection angles are positive. In this case the supersymmetry condition reads\footnote{Here all angles lie in the open interval $(0,1)$.}
\begin{align}
\theta_1 + \theta_2 +\theta_3 =2\,\,.
\end{align}
The lightest state in the NS-sector in that case is given by
\begin{align}
%\psi_{-\frac{1}{2}+ \theta_1} \,\, |\,\theta_{1,2,3}\, \rangle & \hspace{3cm}    M^2 = \frac{1}{2} \left( -\theta_1 + \theta_2 +\theta_3 \right) M^2_s\\
%\psi_{-\frac{1}{2}+ \theta_2} \,\, |\,\theta_{1,2,3}\, \rangle & \hspace{3cm}    M^2 = \frac{1}{2} \left( \theta_1 - \theta_2 +\theta_3 \right)M^2_s\\
%\psi_{-\frac{1}{2}+ \theta_3} \,\, |\,\theta_{1,2,3}\, \rangle  & \hspace{3cm}   M^2 = \frac{1}{2} \left( \theta_1 + \theta_2 -\theta_3 \right)M^2_s\\
\prod^3_{I=1}\psi^I_{-\frac{1}{2}+ \theta_I}\,\, |\,\theta_{1,2,3}\, \rangle^{ab}_{NS}  & \hspace{3cm}
M^2 =\left(1- \frac{1}{2} \left( \theta_1 + \theta_2 +\theta_3 \right)\right) M^2_s\,\,,
\end{align}
which is massless for a supersymmetric configuration.

Given the vertex operator contribution for each complex dimension the corresponding vertex operator takes the form
\begin{align}
\prod^3_{I=1}\psi^I_{-\frac{1}{2}+ \theta_I}\,\, |\,\theta_{1,2,3}\, \rangle^{ab}_{NS}  & ~:~\hspace{1cm} 
V^{(-1)}_{\phi^*} =\Lambda_{ab} \,\, \phi^* e^{-\varphi} \,\, \prod^3_{I =1} \sigma^+_{\theta_I} \, e^{-i(1-\theta_I)H_I } \, \, e^{i k X}~.
\end{align}
It is easy to verify that the conformal dimension of this vertex operator is
%\begin{align}
$h= 2 - \frac{1}{2} \sum^3_{I=1} \theta_I  \, + \,k^2$
%\end{align}
and the state becomes massless ($h=1$) once the supersymmetry condition is satisfied. How do we know that one has to identify this state as the lowest component of an anti-chiral superfield rather than of a chiral superfield? This can be answered by looking at the $U(1)_{WS}$ charge which in the canonical $(-1)$-ghost picture is the same as the $U(1)_R$ charge. In this specific case the $U(1)_{WS}$ charge is
%\begin{align}
$\sum^3_{I=1} (\theta_I-1) =-1$ for the supersymmetric setup, and this should be identified with the scalar of the anti-chiral supermultiplet.
%\end{align}
%Thus this state should combined with a right-handed spinor field to form an anti-chiral supermultiplet. 
The conjugate field is the string going from brane $b$ to $a$ and its vertex operator takes the form (keep in mind that the angles from D6-brane $b$ to D6-brane $a$ are now $-\theta_I$ and thus all negative.)
 \begin{align}
V^{(-1)}_{\phi} =\Lambda_{ba} \,\, \phi_4 \,e^{-\varphi} \,\, \prod^3_{I =1} \sigma^-_{\theta_I} \, e^{i(1-\theta_I) H_I} \, \, e^{i k X}   \,\,.
\label{eq vertex all positive}
 \end{align}
%Below we display the vertex operators for the other three fields (here $J$ indicates the dimension of excitation operator)
%\begin{align}
%\psi_{-\frac{1}{2}+ \theta_J} \,\, |\,\theta_{1,2,3}\, \rangle & ~:~\hspace{1.5cm}
%V^{-1}_{\phi_J} = \Lambda_{ab}\,\phi_J e^{-\varphi} \sigma^+_{\theta_J} e^{-i(1-\theta_J) H_J}  \prod^3_{I \neq J} \sigma^+_{\theta_I} \, e^{i\theta_I H_I} 
%\\
%\psi_{-\frac{1}{2}+ \theta_2} \,\, |\,\theta_{1,2,3}\, \rangle & ~:~\hspace{1.5cm}
%V^{-1}_{\phi_2} =\Lambda_{ab}\, \phi_2 e^{-\varphi} \sigma^+_{\theta_2} e^{-i(1-\theta_2) H_2}  \prod^3_{I \neq 2} \sigma^+_{\theta_I} \, e^{i\theta_I H_I} \\
%\psi_{-\frac{1}{2}+ \theta_3} \,\, |\,\theta_{1,2,3}\, \rangle & ~:~\hspace{1.5cm}
%V^{-1}_{\phi_3} = \Lambda_{ab}\,\phi_3 e^{-\varphi} \sigma^+_{\theta_3} e^{-i(1-\theta_3) H_3}  \prod^3_{I \neq 3} \sigma^+_{\theta_I} \, e^{i\theta_I H_I} \,
\,.
%\end{align}
%Note that in the supersymmetric case they all have $U(1)_{WS}$ charge $+1$ indicating that they belong to chiral super-fields together with their left-handed fermion partners.

The superpartner of $\prod^3_{I=1}\psi_{-\frac{1}{2}+ \theta_I}\,\, |\,\theta_{1,2,3}\, \rangle_{NS}$ is given by the ground state of the R-sector $|\, \theta_{1,2,3}\,\rangle_R$, which is massless independent of the choice of intersection angles and whose vertex operator takes the form
\begin{align}
| \, \theta_{1,2,3} \, \rangle^{ab}_R   ~:~\hspace{1.5cm}   V^{(-1/2)}_{\ov \psi} = \Lambda_{ab} \, \ov \psi_{\dot{\alpha}} e^{-\varphi/2} \, S^{\dot{\alpha}} \prod^3_{I=1} \sigma^+_{\theta_I} e^{i\left(\theta_I -\frac{1}{2} \right) H_I} \,\, e^{i k X}
 \end{align}
 The appearance of the anti-chiral spin field $S^{\dot{\alpha}}$ is dictated by the GSO-projection. Note that the $U(1)_{WS}$ charge
 % \begin{align}
 $\sum^3_{I=1} \left(\theta_I -\frac{1}{2} \right) =\frac{1}{2}$
 %\end{align}
 suggests that this field is identified with a right-handed fermion belonging to an anti-chiral multiplet. The conjugate left-handed fermion is identified with the string going from D6-brane $b$ to D6-brane $a$ and its vertex operator takes the form
\begin{align}
| \, \theta_{1,2,3} \, \rangle^{ba}_R  ~:~\hspace{1.5cm}  V^{(-1/2)}_{\psi} = \Lambda_{ba} \,  \psi_{\alpha} e^{-\varphi/2} \, S^{\alpha} \prod^3_{I=1} \sigma^-_{-\theta_I} e^{i\left(-\theta_I +\frac{1}{2} \right) H_I} \,\, e^{i k X}
\label{eq R-vertex all positive}
\end{align}
Note that the $U(1)_{WS}$ charge for this vertex operator is $-\frac{1}{2}$ indicating that it belongs to a chiral multiplet. This vertex operator is indeed the supersymmetric partner of \eqref{eq vertex all positive} which can be easily checked given that the supercharge is 
\begin{align}
Q^{\alpha} = e^{-\varphi/2} S^{\alpha}  \prod^3_{I=1} \, e^{\frac{i}{2} H_I}\,\,.
\label{eq supercharge}
  \end{align}

Before turning to the second setup let us also display the vertex operators for the states  $\alpha^1_{\theta_1}  \prod^3_{I=1}\psi^I_{-\frac{1}{2}+ \theta_I}\,\, |\,\theta_{1,2,3}\, \rangle $ and
$\left(\alpha^1_{\theta_1} \right)^2  \prod^3_{I=1}\psi^I_{-\frac{1}{2}+ \theta_I}\,\, |\,\theta_{1,2,3}\, \rangle  $
\begin{align*}
\alpha^1_{\theta_1}  \prod^3_{I=1}\psi^I_{-\frac{1}{2}+ \theta_I}\,\, |\,\theta_{1,2,3}\, \rangle^{ba}_R ~:    \hspace{.5cm}   V^{(-1)}_{\Psi_{\tau_1}} = \Lambda_{ba} \, \Psi_{\tau_1}\, e^{-\varphi} \,\, \tau^-_{\theta_1} \, e^{i(1-\theta_1) H_1}\, \prod^3_{I =2} \sigma^-_{\theta_I} \, e^{i(1-\theta_I) H_I} \, \, e^{i k X}   \\
\left(\alpha^1_{\theta_1} \right)^2  \prod^3_{I=1}\psi^I_{-\frac{1}{2}+ \theta_I}\,\, |\,\theta_{1,2,3}\, \rangle^{ba}_R ~:
\hspace{.5cm}   V^{(-1)}_{\Psi_{\omega_1}} = \Lambda_{ba}\, \Psi_{\omega_1}\, e^{-\varphi} \,\, \omega^-_{\theta_I} \, e^{i(1-\theta_1) H_1}\, \prod^3_{I =2} \sigma^-_{\theta_I} \, e^{i(1-\theta_I) H_I} \, \, e^{i k X}
\end{align*}
Again the $U(1)_{WS}$ charge dictates that these are lowest component of chiral super-fields going from brane $b$ to brane $a$.%as indicated by the Chan-Paton matrix. 
The mass of the states are $M^2_{\Psi_{\tau_1}}= \theta_1 M^2_s$ and  $M^2_{\Psi_{\omega_1}}=2 \theta_1 M^2_s$ 
%\begin{align}
%\alpha_{\theta_1} \prod^3_{I=1}\psi_{-\frac{1}{2}+ \theta_I}\,\, |\,\theta_{1,2,3}\, \rangle &  \qquad \qquad \qquad M^2=\theta_1\, M^2_s\\
%\left(\alpha_{\theta_1} \right)^2  \prod^3_{I=1}\psi_{-\frac{1}{2}+ \theta_I}\,\, |\,\theta_{1,2,3}\, \rangle & \qquad \qquad \qquad M^2= 2\theta_1 \, M^2_s
%\end{align}
which can be significantly smaller than the string scale $M_s = 1/\sqrt{\alpha'}$, in case the intersection angle $\theta_1 $ is very small. In section \ref{sec amplitude} we investigate whether and how in such a scenario those light states can be observed.

\subsubsection*{Two positive angles one negative one}
For the sake of concreteness we choose the third angle $\theta_3$ to be negative. The supersymmetry condition is given by
\begin{align}
\theta_1 +\theta_2 + \theta_3 =0 \,\,.
\end{align}
The lightest state  is 
\begin{align}
%\psi_{-\frac{1}{2}+ \theta_1} \,\, |\,\theta_{1,2,3}\, \rangle & \hspace{2cm}    M^2 = \frac{1}{2} \left( -\theta_1 + \theta_2 -\theta_3 \right)  M^2_s\\
%\psi_{-\frac{1}{2}+ \theta_2} \,\, |\,\theta_{1,2,3}\, \rangle & \hspace{2cm}    M^2 = \frac{1}{2} \left( \theta_1 - \theta_2 -\theta_3 \right)  M^2_s\\
\psi^3_{-\frac{1}{2}- \theta_3} \,\, |\,\theta_{1,2,3}\, \rangle^{ab}_{NS}   & \hspace{2cm}   M^2 = \frac{1}{2} \left( \theta_1 + \theta_2 +\theta_3 \right) M^2_s\,\,,
%\psi_{-\frac{1}{2}+ \theta_1}\, \psi_{-\frac{1}{2}+ \theta_2}\, \psi_{-\frac{1}{2}- \theta_3} \,\, |\,\theta_{1,2,3}\, \rangle & \hspace{2cm}
%M^2 =\left(1- \frac{1}{2} \left( \theta_1 + \theta_2 -\theta_3 \right)\right) M^2_s\,\,.
\end{align}
which is massless for a supersymmetric configuration. The corresponding vertex operator is given by
\begin{align}
\psi^3_{-\frac{1}{2}- \theta_3} \,\, |\,\theta_{1,2,3}\, \rangle^{ab}_{NS}  & ~:~\hspace{1cm}  V^{(-1)}_{\phi_3} = \Lambda_{ab}\,\, \phi_3\, e^{-\varphi} \prod^2_{I=1} \sigma^+_{\theta_I} \, e^{i\theta_I H_i} \, \sigma^-_{-\theta_3} \, e^{i (1+\theta_3) H_3} \,\, e^{i k X}
\label{eq vertex 2 positive}
\end{align}
This indeed describes the lowest component of a chiral superfield since the $U(1)_{WS}$ charge is $+1$. Its superpartner is the Ramond ground state $|\, \theta_{1,2,3} \,\rangle_R$ whose vertex operator using table \ref{table Excitations R sector} is given by
\begin{align}
|\, \theta_{1,2,3} \,\rangle^{ab}_R   ~:~\hspace{.5cm} V_{\psi}^{(-1/2)} = \Lambda_{ab} \,  \psi_{\alpha} e^{-\varphi/2} \, S^{\alpha} \prod^2_{I=1} \sigma^+_{\theta_I} e^{i\left(\theta_I -\frac{1}{2} \right) H_I} \,  \sigma^-_{-\theta_3} e^{i\left(\theta_3 +\frac{1}{2} \right) H_3}\, e^{i k X} ~.
\label{eq R-vertex 2 positive}
\end{align}
It is easy to see that the $U(1)_{WS}$ charge is indeed $-\frac{1}{2}$ as expected for a left-handed fermion in a chiral multiplet. Note that applying  the supercharge \eqref{eq supercharge} to this vertex operator one obtains the bosonic vertex operator \eqref{eq vertex 2 positive}.

%\subsubsection*{Two positive angles one negative}
%Let us turn to the second setup where we choose again for the sake of concreteness $\theta_1, \theta_2 \geq 0$ and $\theta_3 \leq 0$ . In that case the vertex operator corresponding to the ground-state $|\, \theta_{1,2,3}\,\rangle $ is given by
%\begin{align}
%|\, \theta_{1,2,3} \,\rangle   ~:~\hspace{.5cm} V_{\psi}^{(-1/2)} = \Lambda_{ab} \,  \psi_{\alpha} e^{-\varphi/2} \, S^{\alpha} \prod^2_{I=1} \sigma^+_{\theta_I} e^{i\left(\theta_I -\frac{1}{2} \right) H_I} \,  \sigma^-_{-\theta_3} e^{i\left(\theta_3 +\frac{1}{2} \right) H_3}\, e^{i k X} ~.
%\label{eq R-vertex 2 positive}
%\end{align}
%It is easy to see that the $U(1)_{WS}$ charge is indeed $-\frac{1}{2}$ as expected for a left-handed fermion in a chiral multiplet. Note that this is also the vertex operator one obtains after applying the supercharge \eqref{eq supercharge} to the bosonic vertex operator \eqref{eq vertex 2 positive}. The corresponding anti-particle is a string stretched from brane $b$ to $a$ and its vertex operator takes the form
%\begin{align}
%V_{\bar\psi}^{(-1/2)} = \Lambda_{ba} \, \ov \psi_{\dot \alpha} e^{-\varphi/2} \, S^{\dot \alpha} \prod^2_{I=1} \sigma^-_{\theta_I} e^{-i\left(\theta_I -\frac{1}%{2} \right) H_I} \,  \sigma^+_{-\theta_3} e^{-i\left(\theta_3 +\frac{1}{2} \right) H_3}\, e^{i k X} \,\,.
%\end{align}

\section{Amplitudes, their factorization and all that
\label{sec amplitude}}
In the previous section we analyzed the configuration of two D6-branes intersecting at non-trivial angles. We gave a recipe for finding the vertex operator corresponding to any physical state. Moreover, we saw that there exists a tower of physical states whose mass is proportional to $M^2\sim \theta M^2_s$, where $\theta$ is the intersection angle in one of the complex dimensions and $M_s$ is the string scale.  If this product is small such states can be light. Here we address the question whether these states can be seen and what their potential signals are.

Before we turn to that issue let us briefly recall the main features of intersecting brane worlds \cite{Blumenhagen:2005mu,Blumenhagen:2006ci,Marchesano:2007de,Bianchi:2009va}. The gauge groups arise from stacks of D6-branes that fill out four- dimensional spacetime and wrap three-cycles in the internal Calabi-Yau threefold. Chiral matter appears at the intersection in the internal space of different cycles wrapped by the D6-brane stacks. The multiplicity of chiral matter between two stacks of D6-branes is given by the topological intersection number of the respective three-cycles.

Many features of a D-brane compactifications, such as chiral matter, gauge symmetry or Yukawa couplings do not crucially depend on the details of the compactification, but rather only on the local structure of the D-brane configurations. Thus it is often times sufficient to investigate a local D-brane setup, described by some quiver theory, and to postpone the embedding into a global setting. This approach is called bottom-up approach and has been initiated in \cite{Antoniadis:2000ena,Aldazabal:2000sa}\footnote{For a systematic search of realistic MSSM D-brane quivers, see \cite{Cvetic:2009yh,Cvetic:2010mm}. For an exhaustive search of global embeddings of such quivers, see \cite{Anastasopoulos:2006da,Anastasopoulos:2010hu}.}.

%Let us present the Madrid-quiver as a concrete example of a promising local D-brane configuration that mimics the MSSM \cite{Ibanez:2001nd,Cremades:2002va,Cremades:2003qj}. It consists of four D-brane stacks giving rise to the gauge symmetry $U(3)_a \times U(2)_b \times U(1)_c \times U(1)_d$. Generically the abelian symmetries are anomalous and get promoted to global symmetries which have to be preserved by perturbative quantities. In the Madrid quiver the linear combination
%\begin{align}
%U(1)_Y= \frac{1}{6} U(1)_a + \frac{1}{2} U(1)_c + \frac{1}{2} U(1)_d
% \end{align}
%remains massless and is identified with the hypercharge. In figure \ref{madridquiver} we display the Madrid quiver with its matter content.

%\begin{figure}
%\begin{center}
%\epsfig{file=MadridQuiver.pdf ,width=80mm}
%\end{center}
%\caption{The Madrid quiver}\label{madridquiver}
%\end{figure}

In the following analysis we have in mind such a local D-brane configuration. However, instead of looking at the whole local configuration we further zoom in and just focus on a subset of the D-brane stacks and investigate the various states localized at the intersection  of two stacks. Let us further specify the setup. We have three stacks of D6-branes wrapping three-cycles on the factorizable six-torus $T^6=T^2\times T^2 \times T^2$ \cite{Aldazabal:2000dg,Ibanez:2001nd, Cremades:2002va}. They intersect each other
non-trivially and give rise to  the following intersection angles\footnote{Any other consistent choice of angles is equally good, but since the CFT computation depends on the concrete form of the vertex operators, we have to make a definite choice of angles.
 }
\begin{align}\nn
\theta^1_{ab}  &> 0  \qquad \qquad  \theta^2_{ab}  >0  \qquad \qquad  \theta^3_{ab}  < 0 \\ \label{eq choice of angles}
\theta^1_{bc}  &> 0  \qquad \qquad  \theta^2_{bc}  > 0  \qquad \qquad  \theta^3_{bc}  <0 \\ \nn
\theta^1_{ca}  &< 0  \qquad \qquad  \theta^2_{ca}  < 0  \qquad \qquad  \theta^3_{ca}  < 0 \,\,.
\end{align}
At each intersection massless chiral fermions appear and, in case of a preserved supersymmetry,
\begin{align}
\theta^1_{ab} + \theta^2_{ab} +\theta^3_{ab} &=0 \qquad \qquad  
\theta^1_{bc} + \theta^2_{bc} +\theta^3_{bc} &=0 \qquad \qquad
\theta^1_{ca} + \theta^2_{ca} +\theta^3_{ca} &=-2
\end{align}
even massless scalars. However we do not always have to enforce them, since the analysis applies independently of whether supersymmetry is preserved or not. Moreover, in the previous section we saw that apart from the massless matter at each intersection there are also massive states whose mass scales with the intersection angle. In scenarios with a low string tension and small intersection angles such states can be fairly light and potentially observed at LHC or future experiments.

Here we compute the scattering amplitude of four chiral fermions
%\begin{align}
$\langle \ov \psi  \,\, \psi \,\, \chi\,\, \ov \chi \rangle$
%\label{eq amplitude trivial}
%\end{align}
where $\psi$ and $\chi$ are the chiral massless fermions localized at the intersection $ab$, and $bc$, respectively. The fields $\ov \psi$ and $\ov \chi$
are their corresponding anti-particle. Let us discuss briefly the naive expectations concerning various limits of this amplitude.

\begin{figure}
\begin{center}
\epsfig{file=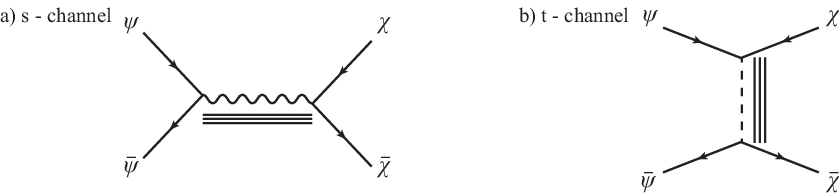,width=145mm}
\end{center}
\caption{The s-channel: the curly line denotes the gauge boson. The t-channel: the dashed line denotes the massless scalar. The solid lines denote massive stringy states.}\label{fig st-channel}
\end{figure}

In the $s$-channel, displayed in figure \ref{fig st-channel}a, one expects the exchange of a gauge boson living on the D-brane stack $b$. Indeed the dominant pole indicates a gauge boson exchange that allows one to normalize the four-point amplitude. Higher poles correspond to exchanges of stringy excitations whose masses scale as $M_s$. Such states can already be observed in the scattering amplitude of four gauge bosons and also in scattering of two fermions onto two gauge bosons. For a sufficiently small string tension, in the $TeV$ range, one may observe signals of these states at LHC \cite{Lust:2008qc,Lust:2009pz}.

On the other hand  in the $t$-channel, displayed in figure \ref{fig st-channel}b, the dominant pole indicates the exchange of a scalar which is massless if supersymmetry is preserved. The latter is a string stretched from D6-brane $a$ to D6-brane $c$. Furthermore one expects additional poles corresponding to exchanges of massive stringy states. In contrast to the $s$-channel exchange particles the masses of those states do not only scale with $M_s$ but also with the intersection angle $\theta_{ac}$. Thus they could be significantly lighter for small intersection angle $\theta_{ac}$ and signals of such states are expected to be observed even before observations of the massive untwisted stringy states.

\subsection{Vertex operators}
For calculating the amplitude $\langle \ov \psi  \,\, \psi \,\, \chi\,\, \ov \chi \rangle$ we need the exact form of the vertex operator. Applying the procedure laid out in section \ref{sec vertex operators} to the choice of intersection angles \eqref{eq choice of angles} one obtains
%With these choices of intersection angles the vertex operators are given by
\begin{align}
ab: \qquad  V^{(-1/2)}_{\psi} = \Lambda_{ab} \psi^{\alpha} \, e^{-\varphi/2} S_{\alpha}  \prod^2_{I=1} \sigma^+_{\theta^I_{ab}}  e^{i \left(\theta^I_{ab} -\frac{1}{2}\right) H_I}\, \sigma^-_{-\theta^3_{ab}}  e^{i \left(\theta^3_{ab} +\frac{1}{2}\right) H_3} \,\, e^{ikX}\,\,.
\end{align}
Its right-handed counterpart is given by
\begin{align}
ba: \qquad  V^{(-1/2)}_{\ov \psi} = \Lambda_{ba} \ov \psi_{\dot{\alpha}} \, e^{-\varphi/2} S^{\dot{\alpha}}  \prod^2_{I=1} \sigma^-_{\theta^I_{ab}}  e^{i \left(-\theta^I_{ab} +\frac{1}{2}\right) H_I}\, \sigma^+_{-\theta^3_{ab}}  e^{i \left(-\theta^3_{ab} -\frac{1}{2}\right) H_3} \,\, e^{ikX}\,\,.
\end{align}
Similarly we get for the $bc$ sector
\begin{align}
bc: \qquad V^{(-1/2)}_{\chi} = \Lambda_{bc} \chi^{\alpha} \, e^{-\varphi/2} S_{\alpha}  \prod^2_{I=1} \sigma^+_{\theta^I_{bc}}  e^{i \left(\theta^I_{bc} -\frac{1}{2}\right) H_I}\, \sigma^-_{-\theta^3_{bc}}  e^{i \left(\theta^3_{bc} +\frac{1}{2}\right) H_3} \,\, e^{ikX}\,\,.
\end{align}
Its right-handed counterpart is given by
\begin{align}
cb: \qquad V^{(-1/2)}_{\ov \chi} = \Lambda_{cb} \ov \chi_{\dot{\alpha}} \, e^{-\varphi/2} S^{\dot{\alpha}}  \prod^2_{I=1} \sigma^-_{\theta^I_{bc}}  e^{i \left(-\theta^I_{bc} +\frac{1}{2}\right) H_I}\, \sigma^+_{-\theta^3_{bc}}  e^{i \left(-\theta^3_{bc} -\frac{1}{2}\right) H_3} \,\, e^{ikX}\,\,.
\end{align}
These vertex operators are sufficient for the amplitude computation $\langle \ov \psi  \,\, \psi \,\, \chi\,\, \ov \chi \rangle$, but before turning to the computation of this amplitude let us also display the vertex operators for the massless scalar\footnote{The scalar is massless only  when supersymmetry is preserved.} as well as for some light massive excitations localized at the intersection of D-branes $a$ and $c$. These will be  the anticipated exchange particles which are related to the dominant and  sub-dominant poles in the t-channel we observe later. Here we assume that the angle $\theta^1_{ca}$ is small, thus the lightest stringy states are generated by exciting with the bosonic operator $\alpha^1_{-\theta^1_{ca}}$.

The vertex operator for the massless scalar $\prod^3_{I=1} \psi^I_{-\frac{1}{2}-\theta^I_{ca}} |\,\theta_{1,2,3}\,\rangle^{ca}_{NS}$ is given by
\begin{align}
V^{(-1)}_{\phi} = \Lambda_{ca} \,\, \phi\,\, e^{-\varphi}   \prod^3_{I=1} \sigma^-_{\theta^I_{ca}} \, e^{i (1+\theta^I_{ca}) H_I} \,\, e^{ik X}
\label{eq vertex phi}
\end{align}
while the one for the first bosonic excitations takes the form
\begin{align}
 V^{(-1)}_{\tilde\phi} = \Lambda_{ca} \, \tilde\phi \, e^{-\varphi} \,\, \widetilde{\tau}^-_{\theta^1_{ca}} \, e^{i(1+\theta^1_{ca}) H_1}\, \prod^3_{I =2} \sigma^-_{\theta^I_{ca}} \, e^{i(1+\theta^I_{ca}) H_I} \, \, e^{i k X}
 \label{eq vertex pi}
\end{align}
which corresponds to the massive state $\alpha^1_{-\theta^1_{ca}} \prod^3_{I=1} \psi^I_{-\frac{1}{2}-\theta^I_{ca}} |\,\theta_{1,2,3}\,\rangle^{ca}_{NS}$ and has mass $M^2 = -\theta^1_{ca} M^2_s$. The second state we consider is $\left(\alpha^1_{-\theta^1_{ca}} \right)^2   \prod^3_{I=1} \psi^I_{-\frac{1}{2}-\theta^I_{ca}} |\,\theta_{1,2,3}\,\rangle^{ca}_{NS} $, that has mass $M^2 =-2 \theta^1_{ca} M^2_s$ and whose vertex operator is given by
 \begin{align}
V^{(-1)}_{\widehat\phi}
= \Lambda_{ca}\, {\widehat\phi} \, e^{-\varphi} \,\, \widetilde{\omega}^-_{\theta^1_{ca}} \, e^{i(1+\theta^1_{ca}) H_1}\, \prod^3_{I =2} \sigma^-_{\theta^I_{ca}} \, e^{i(1+\theta^I_{ca}) H_I} \, \, e^{i k X} \,\,.
\label{eq vertex lambda}
\end{align}
It is easy to check that the conformal dimensions of these vertex operators indeed account for states with mass $M^2= -\theta^1_{ca} M^2_s$ and $M^2= -2 \theta^1_{ca} M^2_s$, respectively.

\subsection{The amplitude}
Given these vertex operators we are now able to compute the amplitude
\begin{align}
{\cal A} = \langle \ov \psi (0) \,\, \psi (x) \,\, \chi(1) \,\, \ov \chi(\infty) \rangle
\end{align}
that allows us to extract the Yukawa coupling between the fields $\psi$, $\chi$ and $\phi$ (as well as $\tilde\phi$ and ${\widehat\phi}$).
Plugging the vertex operators
% one obtains
%\begin{align} \nn
%{\cal A} ~=~& \Tr\left( \Lambda_{ba} \, \Lambda_{ab} \,\Lambda_{bc} \,\Lambda_{cb} \right)  \ov \psi_{\dot{\alpha}} \, \psi^{\alpha}\,  \chi^{\beta}\,  \ov %\chi_{\dot{\beta}}  \int^{1}_0 \,dx  \Big\langle e^{-\varphi/2 (0)  } e^{-\varphi/2 (x)  }
%e^{-\varphi/2 (1)  }
%e^{-\varphi/2 (\infty)  }
%\Big\rangle\\ \nn
%&\times \Big\langle S^{\dot{\alpha}} (0)\,
%S_{\alpha} (x) \, S_{\beta}(1)
%S^{\dot{\beta}} (\infty)\,
%\Big\rangle  \Big\langle e^{ik_1 X(0)} \,  e^{ik_2 X(x)}\,  e^{ik_3 X(1)}\,  e^{ik_4 X(\infty)} \Big\rangle\\
%&\times
%\Big\langle \sigma^+_{-\theta^3_{ab}}(0) \, \sigma^-_{-\theta^3_{ab}}(x)\, \sigma^-_{-\theta^3_{bc}}(1)\,  \sigma^+_{-\theta^3_{bc}}(\infty)\Big\rangle \,\,\,\,\prod^2_{I=1} \Big\langle \sigma^-_{\theta^I_{ab}}(0)\,  \sigma^+_{\theta^I_{ab}}(x)\, \sigma^+_{\theta^I_{bc}}(0)\,  \sigma^-_{\theta^I_{bc}}(\infty) \Big\rangle \nn \\  \nn
%&\times \prod^2_{I=1}
%\Big\langle e^{i\left(-\theta^I_{ab} +\frac{1}{2}\right) H^I(0)}
% e^{i\left(\theta^I_{ab} -\frac{1}{2}\right)H^I(x)}\,  e^{i\left(\theta^I_{bc} -\frac{1}{2}\right) H^I(1)}\, e^{i\left(-\theta^I_{bc} +\frac{1}{2}\right) H^I(\infty)}
%\Big\rangle \\
%&\times \Big\langle e^{i\left(-\theta^3_{ab} -\frac{1}{2}\right) H^3(0)}
% e^{i\left(\theta^3_{ab} +\frac{1}{2}\right)H^3(x)}\,  e^{i\left(\theta^3_{bc} +\frac{1}{2}\right) H^3(1)}\, e^{i\left(-\theta^3_{bc} -\frac{1}{2}\right) H^3(\infty)}
%\Big\rangle \,\,.
%\end{align}
into the correlators given in appendix \ref{app correlators}
 %Combining the various correlators 
 and taking into account the c-ghost contribution $\langle c(0) c(1) c(\infty)  \rangle =x_{\infty}^{-2}$ one obtains for the amplitude
\begin{align} %\nn
{\cal A} &\sim i g_s\,  \Tr\left( \Lambda_{ba} \, \Lambda_{ab} \,\Lambda_{bc} \,\Lambda_{cb} \right)  \ov \psi \cdot \ov \chi \, \psi \cdot  \chi  (2 \pi)^4 \delta^{(4)}\left( \sum^4_i k_i\right) \\
& \hspace{1cm}\times \int^{1}_0 \,dx
\frac{ x^{-1+k_1 \cdot k_2} \left( 1-x\right)^{-\frac{3}{2} +k_2 \cdot k_3}  e^{-S_{cl}(\theta^1_{ab},1-\theta^1_{bc})}\, e^{-S_{cl}(\theta^2_{ab},1-\theta^2_{bc})}\, e^{-S_{cl}(1+\theta^3_{ab},-\theta^3_{bc})} }
 {\left[ I(\theta^1_{ab}, 1-\theta^1_{bc},x) \,  I(\theta^2_{ab}, 1-\theta^2_{bc},x)\,  I(1+\theta^3_{ab}, -\theta^3_{bc},x)  \right]^{\frac{1}{2}}} \,.\nn
\end{align}
Here we used the identification $\sigma^-_{\theta} = \sigma^+_{1+\theta}$ for the bosonic ``twist" and ``anti-twist" fields
%Finally, we also need the correlator containing four bosonic twist fields. It turns out that the following is true 
(see appendix \ref{app bosonic twist fileds} and \cite{Anastasopoulos:2011gn}).
%\begin{align}
%\sigma^-_{\theta}(z) = \sigma^+_{1-\theta}(z) \qquad \qquad  \qquad \sigma^-_{-\theta} (z) = \sigma^+_{1+\theta}(z)
%\end{align}
%This simplifies the computation since one does not have to determine the twist field correlators for different combinations of "twist" and "anti-twist" fields separately but rather can use the result computed for one combination  and appropriately plug in the appropriate angles.

\subsection*{s-channel -- normalization of the amplitude}

Before turning to the t-channel, where we expect the exchange of light stringy states, we will investigate the s-channel which allows us to normalize the amplitude. In order to properly take the limit $x \rightarrow 0 $ we Poiss\'on resum the classical contribution, obtaining
\begin{align}
{\cal A} &\sim i g_s \Tr\left( \Lambda_{ba} \, \Lambda_{ab} \,\Lambda_{bc} \,\Lambda_{cb} \right)  \ov \psi \cdot \ov \chi \, \psi \cdot  \chi  \frac{ (2 \pi)^4 \delta^{(4)}\left( \sum^4_i k_i\right) }{L_{b^1}\, L_{b^2}\, L_{b^3}} \\ &\hspace{.5cm }\times\int^{1}_0 \,dx
\frac{ x^{-1+k_1 \cdot k_2} \left( 1-x\right)^{-\frac{3}{2} +k_2 \cdot k_3} e^{-\widetilde{S}_{cl}(\theta^1_{ab},1-\theta^1_{bc})}\, e^{-\widetilde{S}_{cl}(\theta^2_{ab},1-\theta^2_{bc})}\, e^{-\widetilde{S}_{cl}(1+\theta^3_{ab},-\theta^3_{bc})} }
{ \sqrt{{_2F}_1[\theta^1_{ab},\theta^1_{bc},1;x]\,{_2F}_1[\theta^2_{ab},\theta^2_{bc},1;x]\, {_2F}_1[1+\theta^3_{ab},1+\theta^3_{bc},1;x]
}} \nn \, ,
\end{align}
where $e^{-\widetilde{S}_{cl}} $ in the Hamiltonian form is given by
\begin{align}
e^{-\widetilde{S}_{cl}(\theta,\nu)} =
\prod^3_{i=1}\sum_{p_i, q_i} \exp\left[-\pi
\frac{t(\theta,\nu,x)}{\sin(\pi \theta)}\,\frac{ \alpha'}{ L^2_{b_i} }\,  p^2_i -
\pi \frac{t(\theta,\nu,x) }{\sin(\pi \theta)} \, \frac{ R^2_{x_i}\, R^2_{y_i}}{ \alpha'\,  L^2_{b^i}}
\, q^2_i\right]\,\,.
\end{align}
In the limit $x \rightarrow 0 $ that corresponds to the $s$-channel one has 
\begin{align}
t(\theta,\nu,x)\approx \frac{\sin (\pi\theta)}{ \pi}\left( -\ln(x) + \ln(\delta)\right)
\end{align}
with $\ln (\delta)$ given by
\begin{align}
\ln(\delta) = 2 \psi(1) -\frac{1}{2} \left(\psi(\theta) + \psi(1-\theta) + \psi(\nu) + \psi(1-\nu) \right)\,\,.
\end{align}
Thus the dominant pole in the s-channel is
\begin{align}
{\cal A} =&i g_s\,  {\cal C}  \,\,  \Tr\left( \Lambda_{ba} \, \Lambda_{ab} \,\Lambda_{bc} \,\Lambda_{cb} \right)   (2 \pi)^4 \delta^{(4)}\left( \sum^4_i k_i\right)   \ov \psi \cdot \ov \chi \, \psi \cdot  \chi \nn  \\
&\times \frac{\alpha'^{\frac{3}{2}}}{L_{b^1}\, L_{b^2}\, L_{b^3}}  \int^{0+\epsilon}_0 \,dx \,\,\, x^{-1+s}  \prod^3_{i=1} \sum_{p_i , q_i  } \left(\frac{x}{\delta}\right)^{\frac{\alpha'}{L^2_{b_i}} p^2_i +  \frac{R^2_{x_i} R^2_{y_i}}{\alpha' L^2_{b_i}} q^{2}_i
}~.
\label{eq amplitude s-channel}
\end{align}
For $p_i=q_i =0 $ the amplitude factorizes on the exchange of gauge bosons
\begin{align}
A_4 (k_1,k_2,k_3,k_4) = \mathrm{i}\int \frac{\mathrm{d}^4
k\,\mathrm{d}^4 k^{'}}{(2\pi)^4} \, \,\frac{\sum_{g}
A^g_{\mu} (k_1,k_2,k)A^{g,\mu} (k_3,k_4,k^{'})
\delta^{(4)}(k-k^{'})}{k^2-\mathrm{i} \epsilon } \label{eq unitarity}\,\,.
\end{align}
Knowing the form of the three point amplitude allows us to normalize the amplitude. In eq \eqref{eq unitarity} we sum over all polarizations (vector index $\mu$) and all colors (adjoint index $g$) that can be exchanged.   The three-point amplitude describing the coupling of two fermions to a gauge boson is given by \cite{Cvetic:2006iz}
\begin{align}
A^g_{\mu} (k_1,k_2,k_3)= \mathrm{i} \,g_{D6_b} \,(2\pi)^4
\delta^{(4)}\left(\sum^3_{i=1}k_i\right) \ov{\psi} \sigma^{\mu} \psi\,\,
Tr({\Lambda_{ba} \, \Lambda_{ab}
\Lambda_{bb}})\,\,.
\end{align}
Here $\Lambda_{bb}$ denotes the Chan-Paton matrix of the exchanged gauge boson and the gauge coupling reads\cite{Polchinski:1998rr}
$g^2_{D6_b} = (2 \pi)^4 {\alpha'^{{3}/{2}}} g_s/
{\prod^3_{i=1} 2\pi L_{b_i} }$.  Performing the integral \eqref{eq unitarity} and comparing with \eqref{eq amplitude s-channel} gives for the normalization
%\begin{align}
${\cal C} =2 \pi $,
%\end{align}
where  we used the usual normalization $Tr(\lambda_a\, \lambda_b)=\frac{1}{2} \delta_{ab}$.

Non-vanishing $p_i$ and $q_i$ in \eqref{eq amplitude s-channel} indicate exchanges of KK and winding states, respectively. The exchanges of these states  probe the geometry of the D-brane configuration and thus are very model-dependent. On the other hand there are higher order poles not originating from the world-sheet instanton contributions that are related to stringy excitations. Including sub-dominant terms of the hypergeometric functions in the limit $x \rightarrow 0$ gives
\begin{align}
{\cal A} = &2 i  \pi g_s\,   \,\,  \Tr\left( \Lambda_{ba} \, \Lambda_{ab} \,\Lambda_{bc} \,\Lambda_{cb} \right)   (2 \pi)^4 \delta^{(4)}\left( \sum^4_i k_i\right)   \ov \psi \cdot \ov \chi \, \psi \cdot  \chi \label{eq amplitude s-channel higher} \\
&\times \frac{\alpha'^{\frac{3}{2}}}{L_{b^1}\, L_{b^2}\, L_{b^3}}  \int^{0+\epsilon}_0 \,dx \,\,\, x^{-1+s}  (1+ c_1 x + c_2 x^2 +...)\prod^3_{i=1} \sum_{p_i , q_i  } \left(\frac{x}{\delta}\right)^{\frac{\alpha'}{L^2_{b_i}} p^2_i +  \frac{R^2_{x_i} R^2_{y_i}}{\alpha' L^2_{b_i}} q^{2}_i}~. \nonumber
\end{align}
where $c_i$ are angle dependent coefficients. Note that the sub-dominant poles are integer modded indicating that the mass of the exchanged particles is of order $M_s$, and can be potentially observed at LHC if the string scale is in the $TeV$ range \cite{ArkaniHamed:1998rs,Antoniadis:1998ig}. However the signals are very similar to the ones observed in the scattering of multiple gauge bosons onto  at most two fermions which have been investigated in  \cite{Lust:2008qc,Anchordoqui:2008di,Lust:2009pz,Anchordoqui:2009ja}

\subsection*{t-channel -- exchange of light stringy states}

In this channel we expect the exchange of a massless scalar in case of preserved supersymmetry as well as additional massive states whose mass is basically given by the product of the intersection angle and the string scale $M_s$. If the intersection angle is small these will be long-lived resonances which in case of a low string scale could be observed at LHC.
In addition to these light-stringy excitations one can also observe exchanges of massive stringy states that even in the limit of a vanishing intersection angle remain massive. We will briefly comment on those resonances.

In order to perform this analysis we have to determine the behaviour of $I(\theta,\nu,x)$  and $t(\theta,\nu,x)$ in the limit $x\rightarrow 1$.  Using the properties of the hypergeometric functions displayed in appendix \ref{app hypergeometric} one obtains for $I(\theta,\nu,x)$
\begin{align} \nn
\lim_{x \rightarrow 1} \frac{1}{2\pi}\, I(\theta,\nu,x)
\sim \Gamma_{1-\theta,\nu,1+\theta-\nu} \,\,
%\frac{\Gamma(1-\theta)\,\Gamma(\nu)\,\Gamma(1+\theta-\nu)}{\Gamma(\theta)\,\Gamma(1-\nu)\,\Gamma(\nu-\theta)}\,\,
(1-x)^{\theta-\nu} +  \Gamma_{\theta,1-\nu,1-\theta+\nu} \,\,
%\frac{\Gamma(\theta)\,\Gamma(1-\nu)\,\Gamma(1-\theta+\nu)}{\Gamma(1-\theta)\,\Gamma(\nu)\,\Gamma(\theta-\nu)}\,\,
(1-x)^{\nu-\theta}
\end{align}
where we define $ \Gamma_{\alpha,\beta,\gamma} = \frac{\Gamma(\alpha)\,\Gamma(\beta)\,\Gamma(\gamma)}{\Gamma(1-\alpha)\,\Gamma(1-\beta)\,\Gamma(1-\gamma)} $. For $t(\theta,\nu,x)$ we distinguish among two different scenarios, depending on which angle is larger
\begin{align} \label{eq limits t(x) theta>nu}
\lim_{x\rightarrow 1} t(\theta,\nu,x)&= \frac{\sin(\pi(\theta-\nu) )}{2 \sin(\pi \nu)} \qquad  \qquad \text{for} \qquad \theta > \nu\\
\label{eq limits t(x) theta<nu}
\lim_{x\rightarrow 1} t(\theta,\nu,x) &= \frac{\sin(\pi(\nu-\theta) )}{2 \sin(\pi \nu)} \qquad  \qquad \text{for} \qquad \theta < \nu\,\,.
\end{align}
As a result the amplitude behaves according to
\begin{align*}
{\cal A} =& 2 i \pi g_s\,  \Tr\left( \Lambda_{ba} \, \Lambda_{ab} \,\Lambda_{bc} \,\Lambda_{cb} \right)  \ov \psi \cdot \ov \chi \, \psi \cdot  \chi   (2 \pi)^4 \delta^{(4)}\left( \sum^4_i k_i\right)  \int^{1}_{1-\epsilon} \,dx
 \left( 1-x\right)^{-\frac{3}{2} +k_2 \cdot k_3} \nn \\ \nn
&\times
\left[
\left(
\Gamma_{1-\theta^1_{ab}, 1-\theta^1_{bc}, \theta^1_{ab}+\theta^1_{bc}} (1-x)^{\theta^1_{ab}+\theta^1_{bc}-1} +
\Gamma_{\theta^1_{ab}, \theta^1_{bc}, 2-\theta^1_{ab}-\theta^1_{bc}} (1-x)^{1-\theta^1_{ab}-\theta^1_{bc}}
\right)
\right]^{-\frac{1}{2}}\\ \nn
&\times
\left[
\left(
\Gamma_{1-\theta^2_{ab}, 1-\theta^2_{bc}, \theta^2_{ab}+\theta^2_{bc}} (1-x)^{\theta^2_{ab}+\theta^2_{bc}-1} +
\Gamma_{\theta^2_{ab}, \theta^2_{bc}, 2-\theta^2_{ab}-\theta^2_{bc}} (1-x)^{1-\theta^2_{ab}-\theta^2_{bc}}
\right)
\right]^{-\frac{1}{2}}\\ \nn
&\times
\left[
\left(
\Gamma_{-\theta^3_{ab}, -\theta^3_{bc}, 2+\theta^3_{ab}+\theta^3_{bc}} (1-x)^{1+\theta^3_{ab}+\theta^3_{bc}} +
\Gamma_{1+\theta^3_{ab}, 1+\theta^3_{bc}, -\theta^3_{ab}-\theta^3_{bc}} (1-x)^{-\theta^3_{ab}-\theta^3_{bc}-1}
\right)
\right]^{-\frac{1}{2}} \\
& \times \prod_{p_i,q_i} \prod^2_{i=1} e^{-S^{3}_{cl}(\theta^i_{ab},1-\theta^i_{bc},p_i)  }  e^{-S^{3}_{cl}(\theta^i_{ab},1-\theta^i_{bc},q_i)  } e^{-S^{3}_{cl}(1+\theta^3_{ab},-\theta^3_{bc},p_3)  }  e^{-S^{3}_{cl}(1+\theta^3_{ab},-\theta^3_{bc},q_3)  } \,\,,
\end{align*}
where % $\Gamma_{\alpha,\beta,\gamma}$ is given by
%\begin{align}
%\Gamma_{\alpha,\beta,\gamma} = \frac{\Gamma(\alpha)\,\Gamma(\beta)\,\Gamma(\gamma)}{\Gamma(1-\alpha)\,\Gamma(1-\beta)\,\Gamma(1-\gamma)}
%\end{align}
$e^{-S^{3}_{cl}(\theta,\nu,p) }$ takes the form\cite{Anastasopoulos:2011gn}  \footnote{Recall that all three branes intersect exactly once and for simplicity we assume vanishing Wilson lines and a rectangular torus. With this in mind the intersection angles are given by
\begin{align}
|\sin(\pi \theta^i_{ab})| =  \frac{R_1 R_2}{ L_{a_i} L_{b_i}} \hspace{1cm} |\sin(\pi \theta^i_{bc})| =  \frac{R_1 R_2}{ L_{c_i} L_{a_i}} \hspace{1cm} |\sin(\pi (\theta^i_{ab} -\theta^i_{bc}))| =  \frac{R_1 R_2}{ L_{b_i} L_{c_i}}\,\,.
\label{eq sinus definitions}
\end{align}
For a generalization to setups with non-vanishing Wilson lines and multiple intersections among the three D-branes, see \cite{Cremades:2003qj,Cvetic:2003ch, Abel:2003vv,Cremades:2004wa}.}
\begin{align}
e^{-S^{3}_{cl}(\theta,\nu,p_i) }= \exp\left[- \frac{\pi}{4} \frac{\sin(\pi \theta)\sin(\pi \nu)}{|\sin(\pi (\theta-\nu))|} \frac{L_{b_i}}{\alpha'} p^2_i\right]\,\,.
\end{align}

To simplify the analysis further let us assume that we are in the large volume limit, thus $R_{x_i}, R_{y_i}$ are large. Thus all world-sheet instanton contributions from $p_i, q_i \neq 0$ are negligible. Additionally for the sake of concreteness the intersection angles satisfy
\begin{align}
\theta^1_{ab} +\theta^1_{bc} <1 \qquad   \theta^2_{ab} +\theta^2_{bc} <1 \qquad |\theta^3_{ab} +\theta^3_{bc} |>1.
\label{eq assumption for angles}
 \end{align}

With these assumptions we can pull out the dominant pole and get for the amplitude
\begin{align}
{\cal A} =&     2 i \pi g_s\,  \Tr\left( \Lambda_{ba} \, \Lambda_{ab} \,\Lambda_{bc} \,\Lambda_{cb} \right)  \ov \psi \cdot \ov \chi \, \psi \cdot  \chi   (2 \pi)^4 \delta^{(4)}\left( \sum^4_i k_i\right)
\\ \nn
 & \times
 \int^{1}_{1-\epsilon} \,dx
\frac{ \left( 1-x\right)^{-1 -\frac{1}{2}\sum_I (\theta^I_{ab} +\theta^I_{bc}) +k_2 \cdot k_3}}{ \Gamma^{\frac{1}{2}}_{1-\theta^1_{ab}, 1-\theta^1_{bc}, \theta^1_{ab}+\theta^1_{bc}}\,\Gamma^{\frac{1}{2}}_{1-\theta^2_{ab}, 1-\theta^2_{bc}, \theta^2_{ab}+\theta^2_{bc}}\, \Gamma^{\frac{1}{2}}_{-\theta^3_{ab}, -\theta^3_{bc}, 2+\theta^3_{ab}+\theta^3_{bc}}}
  \\ \nn
& \times  \left[\left( 1+ c_1 (1-x)^{2(1-\theta^1_{ab} -\theta^1_{bc})}\right) \, \left( 1+ c_2 (1-x)^{2(1-\theta^2_{ab} -\theta^2_{bc})}\right)
\left( 1+ c_3 (1-x)^{2(-\theta^3_{ab}-\theta^3_{bc}-1)} \right) \right]^{-\frac{1}{2}}\,\,.
\end{align}
Here the $c_i$'s are given by
\begin{align*}
c_1=\frac{\Gamma_{1-\theta^1_{ab}, 1-\theta^1_{bc}, \theta^1_{ab}+\theta^1_{bc}}}{\Gamma_{\theta^1_{ab}, \theta^1_{bc}, 2-\theta^1_{ab}-\theta^1_{bc}}}\qquad c_2=\frac{\Gamma_{1-\theta^2_{ab}, 1-\theta^2_{bc}, \theta^2_{ab}+\theta^2_{bc}}}{\Gamma_{\theta^2_{ab}, \theta^2_{bc}, 2-\theta^2_{ab}-\theta^2_{bc}}} \qquad c_3=\frac{\Gamma_{-\theta^3_{ab}, -\theta^3_{bc}, 2+\theta^3_{ab}+\theta^3_{bc}}}{\Gamma_{1+\theta^3_{ab}, 1+\theta^3_{bc}, -\theta^3_{ab}-\theta^3_{bc}}}\,\,.
\end{align*}
In the case of preserved supersymmetry ($\sum_I  \theta^I_{ab}  =\sum_I \theta^I_{bc} =0$) 
%\begin{align} \nn
%{\cal A} =&    \ov \psi \cdot \ov \chi \, \psi \cdot  \chi   \int^{1}_{1-\epsilon} \,dx
% \left( 1-x\right)^{-1+k_2 \cdot k_3} \,
%\Gamma^{-\frac{1}{2}}_{1-\theta^1_{ab}, 1-\theta^1_{bc}, \theta^1_{ab}+\theta^1_{bc}}\,\Gamma^{-\frac{1}{2}}_{1-\theta^2_{ab}, 1-\theta^2_{bc}, \theta^2_{ab}+\theta^2_{bc}}\, \Gamma^{-\frac{1}{2}}_{-\theta^3_{ab}, -\theta^3_{bc}, 2+\theta^3_{ab}+\theta^3_{bc}}
%  \\ \nn
%& \times  \left[\left( 1+ c_1 (1-x)^{2(1-\theta^1_{ab} -\theta^1_{bc})}\right) \, \left( 1+ c_2 (1-x)^{2(1-\theta^2_{ab} -\theta^2_{bc})}\right)
%\left( 1+ c_3 (1-x)^{2(-\theta^3_{ab}-\theta^3_{bc}-1)} \right) \right]^{-\frac{1}{2}}
%\end{align}
%The first thing to note is that 
one indeed observes the exchange of a massless scalar \footnote{In the non-susy case the lightest exchange particle has mass $M^2 = \frac{1}{2} \sum^3_{I=1} \left( \theta^I_{ab} + \theta^I_{bc}\right)$.}. This particle is identified with $\phi$ whose vertex operator is displayed in eq. \eqref{eq vertex phi}.

The corresponding physical Yukawa coupling between $\psi$, $\chi$  and $\phi$  is then
\begin{align}
Y_{\psi \chi \phi} \sim    \Gamma^{-\frac{1}{4}}_{1-\theta^1_{ab}, 1-\theta^1_{bc}, \theta^1_{ab}+\theta^1_{bc}}\,\Gamma^{-\frac{1}{4}}_{1-\theta^2_{ab}, 1-\theta^2_{bc}, \theta^2_{ab}+\theta^2_{bc}}\, \Gamma^{-\frac{1}{4}}_{-\theta^3_{ab}, -\theta^3_{bc}, 2+\theta^3_{ab}+\theta^3_{bc}}\,\,.
\label{eq Yukawas}
\end{align}
The angles depend non-holomorphically on the complex structure moduli thus the Gamma-function expressions cannot be part of the holomorphic Yukawa couplings but should rather arise from the K\"ahler potential. The appropriate normalization of the vertex operators going from the string theory basis to the supergravity basis $V^{ST}_{\phi_i} \rightarrow \sqrt{K_{\phi_i\phi_i} } V^{SG}_{\phi_i}$ allows one to extract from \eqref{eq Yukawas} the K\"ahler metrics in complete agreement with 
%
%Thus it is natural to assume that the K\"ahler metrics take the form
% \begin{align}
% {\cal K }_{\psi} &\sim \left(\frac{ \Gamma(\theta^1_{ab})}
%{    \Gamma(1-\theta^1_{ab})}\, \frac{ \Gamma(\theta^2_{ab})}
%{    \Gamma(1-\theta^2_{ab})}\, \frac{ \Gamma(-\theta^3_{ab})}
%{    \Gamma(1-\theta^3_{ab})}\right)^{\frac{1}{2}} \\
% {\cal K }_{\chi} &\sim \left(\frac{ \Gamma(\theta^1_{bc})}
%{    \Gamma(1-\theta^1_{bc})}\, \frac{ \Gamma(\theta^2_{bc})}
%{    \Gamma(1-\theta^2_{bc})}\, \frac{ \Gamma(-\theta^3_{bc})}
%{    \Gamma(1-\theta^3_{bc})}\right)^{\frac{1}{2}} \\
%{\cal K }_{\phi} &\sim \left( \frac{\Gamma(-\theta^1_{ca}) }{   \Gamma(1+\theta^1_{ca}) }
%\frac{\Gamma(-\theta^2_{ca}) }{   \Gamma(1+\theta^2_{ca})} \frac{\Gamma(-\theta^3_{ca}) }{   \Gamma(1+\theta^3_{ca}) }
%\right)^{\frac{1}{2}}
% \end{align}
% which is in complete agreement with 
previous derivations \cite{Lust:2004cx,Bertolini:2005qh,Akerblom:2007np,Honecker:2011sm}.

Let us investigate sub-dominant poles of this amplitude. Recall that we expect massive scalar exchanges, whose mass scales as $M^2 \sim \theta^{I}_{ca} M^2_s$. The expansion $x \rightarrow 1$, including sub-dominant poles gives
 \begin{align*}
& \left[\left( 1+ c_1 (1-x)^{2(1-\theta^1_{ab} -\theta^1_{bc})}\right) \, \left( 1+ c_2 (1-x)^{2(1-\theta^2_{ab} -\theta^2_{bc})}\right)
\left( 1+ c_3 (1-x)^{2(-\theta^3_{ab}-\theta^3_{bc}-1)} \right) \right]^{-\frac{1}{2}}\\
& \hspace{0.5cm}\simeq
1+ c_1 (1-x)^{2(1-\theta^1_{ab} -\theta^1_{bc})} +  c_2 (1-x)^{2(1-\theta^2_{ab} -\theta^2_{bc})} +c_3 (1-x)^{2(-\theta^3_{ab}-\theta^3_{bc}-1)} + ...
 \end{align*}
For concreteness we assume that $1-\theta^1_{ab} -\theta^1_{bc} = -\theta^1_{ca}$ is small and positive. Then the amplitude takes the following form
 \begin{align} 
{\cal A} =&    \ov \psi \cdot \ov \chi \, \psi \cdot  \chi   \int^{1}_{1-\epsilon} \,dx
 \left( 1-x\right)^{-1+k_2 \cdot k_3} \, Y^2_{\psi \chi \phi}
\left( 1+ c_1 (1-x)^{2(1-\theta^1_{ab} -\theta^1_{bc})}+ ...\right),~
\end{align}
the first sub-dominant term suggests that there is a particle with mass $M^2 = -2\theta^1_{ca}>0$  exchanged.

As we have discussed in the beginning of this section, the spectrum in the $ca$ sector indeed reveals a particle with small positive mass $-2\theta^1_{ca} M^2_s$, namely the scalar ${\widehat\phi}$, whose vertex operator is given in eq. \eqref{eq vertex lambda}. Let us stress that there is no coupling to the lightest massive field $\tilde\phi$, which one would have naively expected. This is due to the fact that the  two bosonic twist fields $\sigma$ do not couple to the excited twist field $\tau$, but they only couple to an even excited twist field \cite{Anastasopoulos:2011gn}. In agreement with the latter an inspection of higher poles reveals that the next lightest state exchanged has a mass $-4\theta^1_{ca} M^2_s = 2 M^2 $.\\
A detailed analysis of the next-lighter massive states while straight-forward is beyond the scope of the present investigation. Similarly we do not analyze (higher spin) massive states, whose masses do not vanish for small angles, but we expect similar results as derived in \cite{Lust:2008qc,Anchordoqui:2008di,Lust:2009pz,Anchordoqui:2009ja,Feng:2011qc}. Such an analysis would require a more detailed analysis of the sub-dominant poles of the hypergeometric functions. Note that while signals induced by light stringy states at colliders could be rather difficult to recognize and discriminate from other kinds of Physics Beyond the Standard Model, still these signals are expected to be observed first. Moreover, at higher energy scales one  eventually  will observe higher spin state signatures, which then hint towards a stringy nature.

\section{Summary and Conclusions\label{sec concl}}

%Let us conclude by summarizing our results and drawing some lines for future investigation.

We have carefully studied the spectrum of open strings localized at the intersections of D6-branes. At the cost of being pedantic and partially overlapping with previous investigations \cite{Cvetic:2006iz,Lust:2008qc,Lust:2009pz}, we have identified the ground-states as well as the lowest massive states and displayed the corresponding vertex operators both in the NS- and R-sectors. We had to pay particular attention to the signs of the intersection angles \cite{Abel:2003vv, Cvetic:2003ch, Lust:2008qc, Pesando:2011ce} since the relevant twist fields depend crucially on those. Along the way we provide a dictionary between states and vertex operators for an arbitrary D-brane configuration.
% We have also checked the presence of massless scalars when the angles satisfy supersymmetry preserving conditions. 
We have argued that the masses of the lightest states scale as $M^2_{\theta} \approx \theta M_s^2$ and can thus be parametrically smaller than the string scale if the relevant angle is small. This in turn depends both on the wrapping numbers of the D6-branes and the shape of the tori or orbifolds. We have not address the issue of (supersymmetric) moduli stabilization, which is still open -- at least from a world-sheet CFT vantage point -- and seems to be in tension with chirality. Instead we have considered processes that can expose these light stringy states in their intermediate channel. Relying on previous analysis, we have computed 4-point scattering amplitudes of `twisted' open strings and studied their factorization in the s- and t-channel confirming the presence of the sought for states as sub-dominant poles in the latter. We have found that only evenly excited `twisted' open strings are exchanged in the t-channel, quite differently from what happens for the parent closed-string amplitudes. 

We have not analyzed in any detail the poles corresponding to massive, possibly higher spin, states which remain massive even when some angles are small. Their analysis is tedious and presents significant analogies with the analysis in \cite{Lust:2008qc,Anchordoqui:2008di,Lust:2009pz,Anchordoqui:2009ja}. Notwithstanding the limitations of our analysis, we cannot help drawing some phenomenological conclusions. Assuming a scenario with large extra dimensions and a low scale string tension proves to be realized in Nature, the spectrum of string excitations may be rather `irregular' or at least look very different to the regularly spaced Regge recurrences of the good old Veneziano model. Signals at colliders could be rather difficult to recognize and discriminate from other kinds of Physics Beyond the Standard Model. Yet, the possibility that the lightest massive string excitations be just behind the corner makes worth sharpening our predictions and/or generalizing it to phenomenologically more viable models, possibly including the effect of closed string fluxes and non-perturbative effects.

\section*{Acknowledgements}
We acknowledge M. Cveti{\v c}, F. Fucito, E. Kiritsis, G. Leontaris, J. F. Morales Morera, I. Pesando, G. Pradisi, B. Schellekens, O. Schlotterer, M. Schmidt-Sommerfeld, Ya. Stanev, P. Teresi and T. Weigand for interesting discussions and correspondence. P.~A. is supported by the Austrian Science Fund (FWF) programs P22000, P22114-N16, START project Y435-N16. The work of R.~R. was partly supported by the German Science Foundation (DFG) under the Collaborative Research Center (SFB) 676 â``Particles, Strings and the Early Universeâ".The work of M.~B. was partially supported by the ERC Advanced Grant n.226455 Superfields, by the Italian MIUR-PRIN contract 2009-KHZKRX, by the NATO grant PST.CLG.978785.  P.~A. and R.~R. are grateful to the organizers of the school and workshop of ITN ``UNILHC" PITN-GA-2009-237920 for hospitality during parts of this work. M.~B. and P.~A. would like to thank NORDITA, Stockholm for hospitality during completion of this project.

\newpage
\appendix
\section*{Appendices}
\section{Bosonic twist fields}
\label{app bosonic twist fileds}
Here we display the defining OPE's of the bosonic twist fields discussed in chapter \ref{sec vertex operators}. We start with the bosonic twist fields and then turn to the bosonic anti-twist fields.
\begin{align*}
&\partial Z (z) \, \sigma^+_{\theta} (w) \sim (z-w)^{\theta-1} \tau^+_{\theta} (w)   \qquad \qquad \partial \ov Z (z) \, \sigma^+_{\theta} (w) \sim (z-w)^{-\theta} \widetilde{\tau}^+_{\theta} (w) \\ 
&\partial Z (z) \, \tau^+_{\theta} (w) \sim (z-w)^{\theta-1} \omega^+_{\theta} (w)   \qquad \qquad \partial \ov Z (z) \, \tau^+_{\theta} (w) \sim (z-w)^{-\theta-1} \sigma^+_{\theta} (w) \\ 
&\partial Z (z) \, \omega^+_{\theta} (w) \sim (z-w)^{\theta-1} \rho^+_{\theta} (w)   \qquad \qquad \partial \ov Z (z) \, \omega^+_{\theta} (w) \sim (z-w)^{-\theta-1} \tau^+_{\theta} (w) \\ 
&\partial Z (z) \, \widetilde{\tau}^+_{\theta} (w) \sim (z-w)^{-2+\theta} \sigma^+_{\theta} (w)   \qquad \qquad \partial \ov Z (z) \, \widetilde{\tau}^+_{\theta} (w) \sim (z-w)^{-\theta} \widetilde{\omega}^+_{\theta} (w) \\ 
\nn 
&\partial Z (z) \, \sigma^-_{\theta} (w) \sim (z-w)^{\theta} \tau^-_{\theta} (w)    \qquad \qquad \partial \ov Z (z) \, \sigma^-_{\theta} (w) \sim (z-w)^{-\theta-1} \tau^-_{\theta} (w)\\
&\partial Z (z) \, \tau^-_{\theta} (w) \sim (z-w)^{\theta} \omega^-_{\theta} (w)    \qquad \qquad \partial \ov Z (z) \, \tau^-_{\theta} (w) \sim (z-w)^{-2-\theta} \sigma^-_{\theta} (w)\\
&\partial Z (z) \, \widetilde\tau^-_{\theta} (w) \sim (z-w)^{-\theta-1} \sigma^-_{\theta} (w)    \qquad \qquad \partial \ov Z (z) \, \tau^-_{\theta} (w) \sim (z-w)^{-1+\theta} \widetilde{\sigma}^-_{\theta} (w)\\
&\partial Z (z) \, \widetilde\omega^-_{\theta} (w) \sim (z-w)^{-1+\theta} \widetilde\tau^-_{\theta} (w)    \qquad \qquad \partial \ov Z (z) \, \widetilde\omega^-_{\theta} (w) \sim (z-w)^{-1-\theta} \widetilde{\rho}^-_{\theta} (w)
\end{align*}
The OPE of the bosonic twist and anti-twist fields $\sigma^+_{\theta}$ and $\sigma^-_{\theta}$, whose conformal dimensions are $h_{\sigma^+_{\theta}}= \frac{1}{2}\theta\left(1-\theta \right)$ and  $h_{\sigma^-_{\theta}}= -\frac{1}{2}\theta\left(1+\theta \right)$,   with the conformal fields $\partial Z$ and $\partial \ov Z$ suggest the following identification
\begin{align}
\sigma^-_{\theta} = \sigma^+_{1+\theta}\,\,.
\end{align}
which %This identification of twist and anti-twist fields 
can be  easily generalized to excited twist fields \cite{Anastasopoulos:2011gn}.
With these OPE's one can determine the conformal dimension of the respective twist fields. We summarize our findings in table \ref{table conformal dimensions}.

%{\bf MB MAYBE A SIMPLE LIST OR EQUATION  WITH BIGGER LETTERS AND SYMBOLS WOULD BE BETTER READABLE!!! }

%\begin{table}[h] \centering
%{\tiny
%\begin{tabular}{|c|c|c|c|c|c|c|c|c|} \hline
%\text{Fields} & $\sigma^+_{\theta}$ &$\tau^+_{\theta}$ & 
%$\omega^+_{\theta}$ &$\widetilde{\tau}^+_{\theta}$ & $\sigma^-_{\theta}$ & $\tau^-_{\theta}$ &$\widetilde{\tau}^-_{\theta}$ & $\widetilde{\omega}^-_{\theta}$\\ \hline
%\text{conf. dim.} & $\frac{1}{2}\theta(1-\theta)$& $\frac{1}{2}\theta(3-\theta)$ & $\frac{1}{2}\theta(5-\theta)$ & $\frac{1}{2}(\theta+2)(1-\theta)$ & $-\frac{1}{2} \theta(1+\theta) $&$\frac{1}{2}( 2-\theta) (1+\theta)$ & $ -\frac{1}{2} \theta(3+\theta)$ & $-\frac{1}{2} \theta(5+\theta) $\\\hline
%\end{array}
%\end{tabular}
%}
%\caption{\small {The conformal dimensions of bosonic twist fields.}} % \vspace{3mm}
%\label{table conformal dimensions}
%\end{table}

\begin{table}[h] \centering
\begin{tabular}{| l | l || l | l |}
\hline
\multicolumn{2}{|c||}{Positive angles}&
\multicolumn{2}{|c|}{Negative angles} 
\\
\hline
\hline
Fields & conf. dim. & Fields & conf. dim. \\ \hline \hline
$\sigma^+_{\theta}$    &   $\frac{1}{2}\theta(1-\theta)$   &   $\sigma^-_{\theta}$    &    $-\frac{1}{2} \theta(1+\theta) $    \\
$\tau^+_{\theta}$    &   $\frac{1}{2}\theta(3-\theta)$   &  $\tau^-_{\theta}$    &   $\frac{1}{2}( 2-\theta) (1+\theta)$  ~~~    \\
$\omega^+_{\theta}$    &   $\frac{1}{2}\theta(5-\theta)$   &   $\widetilde{\tau}^-_{\theta}$   &    $ -\frac{1}{2} \theta(3+\theta)$    \\
$\widetilde{\tau}^+_{\theta}$    &   $\frac{1}{2}(\theta+2)(1-\theta)~~~$   &   $\widetilde{\omega}^-_{\theta}$   &    $-\frac{1}{2} \theta(5+\theta) $    \\
\hline
\end{tabular}\nn
\caption{\small {The conformal dimensions of bosonic twist fields.}} % \vspace{3mm}
\label{table conformal dimensions}
\end{table}

\section{Massive states}
\label{app massive states}
In this appendix we  discuss various other massive states localized at the intersection of two D-branes. We apply the dictionary laid out in chapter \ref{sec vertex operators}  and display their corresponding vertex operators. 
\subsection*{Postive angles}
The lowest fermionic excitations in the NS-sector are given by \cite{Aldazabal:2000dg}
\begin{align}
\psi^I_{-\frac{1}{2}+ \theta_I} \,\, |\,\theta_{1,2,3}\, \rangle^{ab}_{NS} & \hspace{3cm}    M^2 = \frac{1}{2} \left( -\theta_I + \sum_{J\neq I} \theta_J  \right) M^2_s \label{eq mass boson}
%\psi_{-\frac{1}{2}+ \theta_2} \,\, |\,\theta_{1,2,3}\, \rangle^{ab}_{NS} & \hspace{3cm}    M^2 = \frac{1}{2} \left( \theta_1 - \theta_2 +\theta_3 \right)M^2_s\\
%\psi_{-\frac{1}{2}+ \theta_3} \,\, |\,\theta_{1,2,3}\, \rangle^{ab}_{NS}  & \hspace{3cm}   M^2 = \frac{1}{2} \left( \theta_1 + \theta_2 -\theta_3 \right)M^2_s
%\prod^3_{I=1}\psi_{-\frac{1}{2}+ \theta_I}\,\, |\,\theta_{1,2,3}\, \rangle^{ab}_{NS}  & \hspace{3cm}
%M^2 =\left(1- \frac{1}{2} \left( \theta_1 + \theta_2 +\theta_3 \right)\right) M^2_s\,\,,
\end{align}
whereas the corresponding vertex operators take the form
\begin{align*}
\psi^I_{-\frac{1}{2}+ \theta_I} \,\, |\,\theta_{1,2,3}\, \rangle^{ab}_{NS}  & ~:~\hspace{1.5cm}
V^{(-1)}_{\Phi_I} = \Lambda_{ab}\,\Phi_I e^{-\varphi} \sigma^+_{\theta_I} e^{i(\theta_I-1) H_I}  \prod^3_{J \neq I} \sigma^+_{\theta_J} \, e^{i\theta_J H_J} \, e^{ikX}
%\\
%\psi_{-\frac{1}{2}+ \theta_2} \,\, |\,\theta_{1,2,3}\, \rangle^{ab}_{NS}  & ~:~\hspace{1.5cm}
%V^{-1}_{\phi_2} =\Lambda_{ab}\, \phi_2 e^{-\varphi} \sigma^+_{\theta_2} e^{-i(1-\theta_2) H_2}  \prod^3_{I \neq 2} \sigma^+_{\theta_I} \, e^{i\theta_I H_I} \\
%\psi_{-\frac{1}{2}+ \theta_3} \,\, |\,\theta_{1,2,3}\, \rangle^{ab}_{NS}  & ~:~\hspace{1.5cm}
%V^{-1}_{\phi_3} = \Lambda_{ab}\,\phi_3 e^{-\varphi} \sigma^+_{\theta_3} e^{-i(1-\theta_3) H_3}  \prod^3_{I \neq 3} \sigma^+_{\theta_I} \, e^{i\theta_I H_I} \,
\,.
\end{align*}
Their corresponding superpartners are given by the following excitation of the R-groundstate 
\begin{align*} \psi^I_{-1+\theta_I} |\,\theta_{1,2,3}\,  \rangle^{ab}_{R}  & ~:~\hspace{1.5cm}
V^{(-1/2)}_{\psi_I} = \Lambda_{ab}\,\psi^{\alpha}_I \,S_{\alpha} e^{-\varphi/2} \sigma^+_{\theta_I} e^{i\left(\theta_I-\frac{3}{2}\right) H_I}  \prod^3_{J \neq I} \sigma^+_{\theta_J} \, e^{i\left(\theta_J -\frac{1}{2}\right) H_J} \, e^{ikX}
\end{align*} 
where we applied  tables \ref{table Excitations NS sector} and \ref{table Excitations R sector}  for the vertex operators. Their masses are given by $M^2 =  \left(1-\theta_I\right) M^2_s$, which coincides with the bosonic masses \eqref{eq mass boson} when supersymmtry is preserved. Via the same procedure we can get the vertex operator for the state $\alpha^I_{-\theta_I}\psi^I_{-\frac{1}{2}+ \theta_I} \,\, |\,\theta_{1,2,3}\, \rangle^{ab}_{NS}$ and its superpartner $\alpha^I_{-\theta_I} \psi^I_{-1+\theta_I} |\,\theta_{1,2,3}\,  \rangle^{ab}_{R}$ 
\begin{align*}
\alpha^I_{-\theta_I}\psi^I_{-\frac{1}{2}+ \theta_I} \,\, |\,\theta_{1,2,3}\, \rangle^{ab}_{NS} & ~:~\hspace{1cm}
V^{(-1)}_{\widetilde{\Phi}_I} = \Lambda_{ab}\,\widetilde{\Phi}_I e^{-\varphi} \tau^+_{\theta_I} e^{i(\theta_I-1) H_I}  \prod^3_{J \neq I} \sigma^+_{\theta_J} \, e^{i\theta_J H_J} \, e^{ikX}
\\
\alpha^I_{-\theta_I}\psi^I_{-1+\theta_I} |\,\theta_{1,2,3}\,  \rangle^{ab}_{R} & ~:~\hspace{1cm}
V^{(-1/2)}_{\widetilde\psi_I} = \Lambda_{ab}\,\widetilde{\psi}^{\alpha}_I \,S_{\alpha} e^{-\varphi/2} \sigma^+_{\theta_I} e^{i\left(\theta_I-\frac{3}{2}\right) H_I}  \prod^3_{J \neq I} \sigma^+_{\theta_J} \, e^{i\left(\theta_J -\frac{1}{2}\right) H_J} \, e^{ikX}
\end{align*}
whose masses are given by $M^2_{\widetilde{\Phi}}= \sum_{I\neq J} \frac{\theta_J}{2}$ and $M^2_{\widetilde\psi_I}=M^2_s$ which as expected coincide for preserved supersymmetry. 
%\section{The twist and anti-twist fields $\sigma^+_{\theta}$ and $\sigma^-_{\theta}$
%\label{app twists} }
%Let us take a look at the OPE of these two fields with the conformal fields $\partial Z^I$ and $\partial \ov Z^I$. As shown in section \ref{sec NS sector}
%we have the following OPE's
%\begin{align}
%&\partial Z^I (z) \, \sigma^+_{\theta} (w) \sim (z-w)^{\theta-1} \tau^+_{\theta} (w)   \qquad \qquad \partial \ov Z^I (z) \, \sigma^+_{\theta} (w) \sim (z-w)^{-\theta} \widetilde{\tau}^+_{\theta} (w) \\ \nn \\
%&\partial Z^I (z) \, \sigma^-_{\theta} (w) \sim (z-w)^{\theta} \widetilde{\tau}^-_{\theta} (w)    \qquad \qquad \partial \ov Z^I (z) \, \sigma^-_{\theta} (w) \sim (z-w)^{\theta-1} \tau^-_{\theta} (w)
%\end{align}
%Their conformal dimensions are given by
%\begin{align}
%$h_{\sigma^+_{\theta}}= \frac{1}{2}\theta\left(1-\theta \right)$ and  $h_{\sigma^-_{\theta}}= -\frac{1}{2}\theta\left(1+\theta \right)$
%\end{align}
%which suggests the following identification
%\begin{align}
%\sigma^-_{\theta} = \sigma^+_{1+\theta}\,\,.
%\end{align}
%Analogously one can also show that
%\begin{align}
%\sigma^-_{-\theta} = \sigma^+_{1-\theta}\,\,.
%\end{align}
%For more details on these identifications, specifically in the context of excited twist fields, see \cite{Anastasopoulos:2011gn}.

\section{Correlators
\label{app correlators}}
Below we display the necessary correlators for the computation of the four point amplitude considered in section  \ref{sec amplitude}.
\begin{align}
\Big\langle e^{-\varphi/2 (0)  } e^{-\varphi/2 (x)  }
e^{-\varphi/2 (1)  }
e^{-\varphi/2 (\infty)  }
\Big\rangle = \left[x (1-x)\right]^{-\frac{1}{4}} \, x^{-\frac{3}{4}}_{\infty} \\
%\end{align}
%\begin{align}
\Big\langle S^{\dot{\alpha}} (0)\,
S_{\alpha} (x) \, S_{\beta}(1)
S^{\dot{\beta}} (\infty)\,
\Big\rangle =\epsilon_{\alpha \beta} \, \epsilon^{\dot{\alpha} \dot{\beta}} \, \left(1-x \right)^{-\frac{1}{2}} \, x_{\infty}^{-\frac{1}{2}} \\
%\end{align}
%\begin{align}
\Big\langle e^{ik_1 X(0)} \,  e^{ik_2 X(x)}\,  e^{ik_3 X(1)}\,  e^{ik_4 X(\infty)} \Big\rangle = x^{k_1\cdot k_2} \, (1-x)^{k_2 \cdot k_3} \,  \, x_{\infty}^{k_4(k_1 +k_2 +k_3)} \\
%\end{align}
%\begin{align}
\Big\langle
e^{i\alpha H^I(0)}\,  e^{i\beta H^I(x)}\, e^{i\gamma H^I(1)}\, e^{i\delta H^I(\infty)} \Big \rangle = x^{\alpha \, \beta} (1-x)^{\beta \, \gamma} x^{\delta(\alpha +\beta+ \gamma)}_{\infty}\,\,.
\end{align}
%Finally, we also need the correlator containing four bosonic twist fields. It turns out that the following is true (see appendix \ref{app twists} and %\cite{Anastasopoulos:2011gn})
%\begin{align}
%\sigma^-_{\theta}(z) = \sigma^+_{1-\theta}(z) \qquad \qquad  \qquad \sigma^-_{-\theta} (z) = \sigma^+_{1+\theta}(z)
%\end{align}
%This simplifies the computation since one does not have to determine the twist field correlators for different combinations of "twist" and "anti-twist" fields separately but rather can use the result computed for one combination  and appropriately plug in the appropriate angles.
For the bosonic  twist field correlator one finds \cite{Cvetic:2003ch,Abel:2003vv,Lust:2004cx,Lust:2008qc,Anastasopoulos:2011gn}
\begin{align*}
x^{\nu(1-\nu)}_{\infty}\,\,\langle \sigma^+_{1-\theta}(0)\, \sigma^+_{\theta}(x) \,
\sigma^+_{1-\nu}(1) \, \sigma^+_{\nu} (\infty)\rangle =
x^{-\theta(1-\theta)} \left(1-x\right)^{-\frac{1}{2} (\theta + \nu)
+\theta\nu} I^{-\frac{1}{2}}( \theta, \, \nu, \, x) e^{-S_{cl}(\theta,\nu)} \,\,.
\end{align*}
Here $I(x,\theta, \nu)$ is given by
\begin{align*}
I(\theta,\nu,x)= \frac{1}{2\pi} \big[B_1(\theta,\nu) \,{G_2}(x)
H_1(1-x)+ B_2(\theta,\nu) \, G_1(x) {H_2}(1-x)\big]\,\,,
\end{align*}
where
\begin{align*}
\begin{gathered}
B_1(\theta,\nu)=\frac{\Gamma(\theta)\,\Gamma(1-\nu)}{\Gamma(1+\theta-\nu)}\qquad
B_2(\theta,\nu)=\frac{\Gamma(\nu)\,\Gamma(1-\theta)}{\Gamma(1+\nu-\theta)}\\
G_1(x)= {_2F}_1[\theta,1-\nu,1;x]\qquad
G_2(x)= {_2F}_1[1-\theta,\nu,1;x]\\
H_1(x)= {_2F}_1[\theta,1-\nu,1+\theta-\nu;x]\qquad
 H_2(x)={_2F}_1[1-\theta,\nu,1-\theta+\nu;x]\,\,.
 \end{gathered}
\end{align*}
The classical contribution takes the (Lagrangian) form\footnote{For the sake of clarity here we simplify the configuration by assuming that all three D-branes are intersecting exactly once and all Wilson lines are vanishing. A generalization of the results can be easily obtained using the results of \cite{Cremades:2003qj,Abel:2003vv,Cremades:2004wa,Lust:2008qc}}
\begin{align}
e^{-S_{cl}(\theta,\nu)}=\sum_{\widetilde{p}_i, q_i} \exp\left[-\pi \frac{\sin(\pi \theta) } {t(\theta,\nu,x)} \,\frac{L^2_{b^i}}{\alpha'} \,  \widetilde{p}_i^2- \pi \frac{t(\theta,\nu,x)} {\sin(\pi \theta)} \, \frac{ \, R^2_{x_i}\, R^2_{y_i}} {\alpha' L^2_{b^i}}  \, q^2_i\right]
\end{align}
with $t(\theta,\nu,x)$ given by
\begin{align}
t(\theta,\nu,x)=
 \frac{\sin(\pi \theta)}{2 \pi} \left( \frac{B_1 H_1(1-x)}{G_1(x)} + \frac{B_2 H_2(1-x)}{G_2(x) }
 \right)\,\,
 \end{align}
 and Here $R_{x_i}$ and  $R_{y_i}$ are the radii of the two torus and $L_a$ and $L_b$ denotes the length of the brane $a$ and $b$, respectively.

 \section{Properties of hypergeometric functions \label{app hypergeometric}}
In this appendix we display various properties of hypergeometric functions that we will use throughout the paper. The hypergeometric function is given by
\begin{align}
{_2F}_1[\theta,1-\nu,1,z] = \frac{1}{\Gamma(\theta) \, \Gamma(1-\nu)} \sum^{\infty}_{n=0} \frac{\Gamma(\theta+n) \, \Gamma(1-\nu+n)}{\Gamma(n)}  \frac{z^n}{n !}\,\,.
\end{align}
where the series is only convergent for $|z|\leq 1$. Below we display some relations of the hypergeometric functions, starting with
\begin{align}
{_2F}_1[a,b,c,z]=(1-z)^{c-a-b}{_2F}_1c-a,c-b,c,z]\,\,.
\end{align}
For $a+b-c \neq m$, where $m \in \mathbf{Z}$
\begin{align}
{_2F}_1[a,b,c,z]&=\frac{\Gamma(c)\Gamma(c-a-b)}{\Gamma(c-a)\Gamma(c-b)} {_2F}_1[a,b,a+b-c+1,1-z] \\ &\hspace{1cm}
(1-z)^{c-a-b} \, \frac{\Gamma(c)\Gamma(a+b-c}{\Gamma(a)\Gamma(b)} {_2F}_1[c-a,c-b,c-a-b+1,1-z] \,\,.\nn
\end{align}
For $c=a+b$ one obtains
\begin{align}
{_2F}_1[a,b,a+b,z]&=\frac{\Gamma(a+b)}{\Gamma(a)\Gamma(b)}  \sum^{\infty}_{n=0}   \frac{(a)_n (b)_n}{(n!)^2}   \\
& \hspace{1cm }\times \left[ 2\psi(n+1) -\psi(a+n)-\psi(b+n)-\ln(1-z)\right](1-z)^n\,\,, \nn
\end{align}
where $\psi(z)$ is the Digamma function $\psi(z)= \frac{d \ln \Gamma(z)}{dz}$ and $(a)_n$ denotes Pochhammer's symbol $(a)_n= \frac{\Gamma(a+n)}{\Gamma(a)}$.

%\clearpage \nocite{*}
%\bibliographystyle{JHEP}
%\bibliography{reference.bib}
%\end{document}

\clearpage \nocite{*}
\providecommand{\href}[2]{#2}\begingroup\raggedright\endgroup

\end{document}